\documentclass[amsmath,comment,amssymb,twocolumn,showpacs,prb,aps]{revtex4}
\usepackage{graphicx}
\usepackage{amssymb}
\usepackage{epsfig}
\usepackage{dcolumn}% Align table columns on decimal point
\usepackage{bm}% bold math

\newcommand{\be}{\begin{equation}}
\newcommand{\ee}{\end{equation}}
\newcommand{\bea}{\begin{eqnarray}}
\newcommand{\eea}{\end{eqnarray}}

\newcommand{\nn}{\nonumber }

\begin{document}
\title{Localization of disordered bosons and magnets in random fields}

\author{Xiaoquan Yu${}^{1,2}$}
\author{Markus M\"uller${}^{3}$}
\affiliation{ ${\ }^1$International School for Advanced Studies (SISSA), via Bonomea 265, 34136 Trieste, Italy \\
${\ }^2$New Zealand Institute for Advanced Study, Centre for Theoretical Chemistry and Physics,
Massey University, Auckland 0745, New Zealand\\
${\ }^3$The Abdus Salam International Center for Theoretical Physics, Strada Costiera 11, 34151 Trieste, Italy}

\pacs{61.43.-j, 67.25.dj}

\date{\today }

\begin{abstract}
We study localization properties of disordered bosons and spins in random fields at zero temperature. We focus on two representatives of different symmetry classes,
 hard-core bosons (XY magnets) and Ising magnets in random transverse fields, and contrast their physical properties. We describe localization properties using a locator expansion on general lattices. For 1d Ising chains, we find non-analytic behavior of the localization length as a function of energy at $\omega=0$, $\xi^{-1}(\omega)=\xi^{-1}(0)+A|\omega|^\alpha$, with $\alpha$ vanishing at criticality. This contrasts with the much smoother behavior predicted for XY magnets.  
We use these results to approach the ordering transition
on Bethe lattices of large connectivity $K$, which mimic the limit of high dimensionality. In both models, in the paramagnetic phase with uniform disorder, the localization length is found to have a local maximum at $\omega=0$. For the Ising model, we find activated scaling at the phase transition, in agreement with infinite randomness studies. In the Ising model long range order is found to arise due to a delocalization and condensation initiated at $\omega=0$, without a closing mobility gap.  We find that Ising systems establish order on much sparser (fractal) subgraphs than XY models. Possible implications of these results for finite-dimensional systems are discussed.  
\end{abstract}

\maketitle

%%%%%%%%%%%%%%%%%%%%%%%%%%%%%%%%%%%%%%%%%%%%%%%%%%%%5

%%%%%%%%%%%%%%%%%%%%%%%%%%%%%%%%%%%%%%%%%%%%%%%%%%%

\section{Introduction}

The localization properties of excitations in disordered, interacting many body systems attract 
interest from many different communities. This issue is not only relevant for the quantum transport of 
electrons,~\cite{AndersonFleishman,BerkovitsShklovskii,Mirlin,
Altshuler,HuseOganesyan} or spins,~\cite{Prelovsek,PalHuse,Buccheri,deLuca} but also for the problem of noise protection 
in solid state quantum computation, the protection of topological order~\cite{HuseSondhi1304} or the dynamics in ultracold atoms 
subject to disorder potentials.~\cite{Damski,Inguscio,Aspect} Disordered magnets and bosons are among the simplest 
and most promising many body systems, both experimentally and theoretically, 
to study the arising conceptual questions regarding the interplay of disorder, 
interactions and the formation of long range order. In the present paper we address localization properties of the quantum disordered phase of
such systems, and study how they approach the quantum phase transition to the ordered phase.

As compared to interacting systems, where many questions remain open, the localization of 
non-interacting quantum particles in a random potential have been studied rather extensively and are well understood.
In high enough dimensions, 
single particle wavefunctions undergo the 
Anderson localization transition from a metallic to an insulating phase upon increasing 
the disorder.~\cite{Anderson} The behavior of such systems in the presence of
 interactions is a much more involved subject. Recently, the long-standing question as to
 the stability of Anderson insulators with respect to weak interactions has attracted renewed interest.
 Both analytical arguments and numerics suggest that \textquotedblleft many body localized\textquotedblright 
~phases with no intrinsic diffusion and transport exist in the presence of strong enough disorder,
 if the interactions are sufficiently short ranged and weak.~\cite{AndersonFleishman,Altshuler,Mirlin} 
It has been predicted, that as a function of various control parameters, such as increasing interaction
 strength, decreasing disorder and or increasing temperature, these systems may undergo a delocalization transition to an intrinsically 
conducting phase, which does not rely anymore on an external bath to sustain transport.~\cite{BerkovitsShklovskii, Altshuler, HuseOganesyan, PalHuse, Prelovsek}

Until recently, the localization properties in disordered bosonic systems have received less attention than those
 of fermions, even though Anderson's original work on localization was actually motivated 
by the apparent absence of diffusion in spin systems.
However, the recent realizations of disordered bosons in optical 
lattices~\cite{Damski,Inguscio,Aspect} and spin ladders~\cite{Ruegg} have spurred 
renewed interest in questions regarding the localization of bosons 
in disorder.~\cite{GiamarchiSchulz, Prokofev, Shklovskii, SanchezPalencia, Kollath, MuellerShklovskii, Nattermann, Roscilde, Zamponi, GingrasMelko,YuMueller2012,Shapiro} Some of these issues have previously arisen in the context of dirty superconductors~\cite{dirtysuperconductor,MaLee,Goldmanreview} or in studies of $^{4}$He in porous media,~\cite{Cao,Reppy,Davis} and have recently gained a much wider range of applicability. 

In all the above mentioned systems, interactions between the bosons are essential to prevent a collapse into
the lowest-lying single particle eigenstate. Consequently, the problem of bosonic localization is inherently 
an interacting problem, which requires a many body approach from the outset. 

An important feature that distinguishes  bosons from (repulsive) fermions, 
is their ability to condense into a superfluid state with long range order and perfect, dissipation less transport. 
Nevertheless, when subjected to too strong disorder, global phase coherence is suppressed 
and the bosons localize into an insulating state, 
the so-called \textquotedblleft Bose glass\textquotedblright.~\cite{FisherBoseglass, GiamarchiSchulz, 3ddisoerderedhardcorebosons}
While it seems intuitive to consider the disorder-driven quantum phase transition 
from Bose glass to superfluid as a kind of \textquotedblleft collective boson delocalization\textquotedblright,~the precise relation between this phenomenon and single particle Anderson transition is not well understood.~\cite{HertzAnderson} Certain qualitative features might well carry over from the single particle case, but one should also expect significant differences due to the statistics of the particles and the incipient long range order.~\cite{positivemagnetoresistance}

In this article we address these questions by developping a perturbative technique which analyzes localization properties
in strongly disordered bosonic systems. 
A short account of part of these results has been presented in Ref.~\onlinecite{positivemagnetoresistance}. 
The present study is complementary to the analysis of bosonic excitations {\em within} long-range ordered, 
but strongly inhomogeneous phases,~\cite{ChalkerGurarie, GunnLee, CordMueller, MonthusGarelphonons, AlvarezLaflorencie} where a substantial amount 
of literature has discussed the localization properties of Goldstone modes, spin waves and phonons at low energies. 
Here, we focus instead on understanding the {\em insulating}, quantum disordered phase of random bosonic systems. 
To this end we consider two prototypical boson and spin models in random fields, and study the difference between discrete (Ising) 
and continuous (XY) symmetry. At the same time we will revisit and correct the recent approach by Ioffe,
M\'ezard and Feigel'man~\cite{IoffeMezard,FeigelmanIoffeMezard} to these questions.

This paper is organized as follows: In Sec.~\ref{sec:models}, we introduce the canonical bosonic and spin models we are going to study, and contrast them with Anderson's model of non-interacting, disordered fermions. In Sec.~\ref{sec:review} we briefly review existing approaches to disordered bosonic systems and summarize their key findings, before giving an overview of our results. 
In Sec.~\ref{s:decay} we use the locator expansion introduced in Ref.~\onlinecite{positivemagnetoresistance} to calculate the decay rate of local excitations to leading order in the hopping or exchange. This allows us to characterize a localization length of these many body excitations. 
Sec.~\ref{s:1d} benchmarks the results of the leading order expansion in the exchange against the exactly solvable 1d  Ising chain in a random transverse field. We show that to leading order the localization length of the Jordan-Wigner (JW) fermions agrees with the locator expansion for spin excitations, and  that due to the chiral symmetry this result is actually exact to all orders at $\omega=0$. The localization length is found to decrease non-analytically with increasing energy ($\omega>0$) of the JW fermions. In Sec.~\ref{s:delocalization} we analyze the boson and spin models on a Cayley tree (Bethe lattice), as a way to approach localization phenomena in high dimensions. At large connectivity, the low energy excitations are well described by taking into account the most relevant subleading corrections to the leading order expansion in the exchange.  This allows us to study the approach to the ordering transition (bosonic delocalization). 
For uniformly distributed disorder we show that all low energy excitations remain localized in the quantum disordered phase in both models under study. This implies that order sets in by a delocalization at $\omega=0$ and not by a collapsing mobility edge.
We also show that the spin symmetry affects the nature of the transition significantly: Ising systems exhibit infinite randomness characteristics, while XY models show common power law scalings for low energy excitations.   
The possible implications for finite dimensions and open questions are discussed in Sec.~\ref{s:discussion}. A confirmation of the locator expansion by simple perturbation theory is relegated to an appendix.

\section{Models}
\label{sec:models}
\subsection{Fermions versus hardcore bosons}
The phenomenon of Anderson localization is well epitomized by the model of a spinless quantum particle hopping on a lattice,~\cite{Anderson} as it arises, e.g., in the impurity band of a semiconductor once interactions are neglected:
\bea
\label{Anderson}
H=-\sum_{i}\epsilon_{i}n_{i}-\sum_{\langle i,j \rangle}t_{ij}\left(c^{\dag}_{i}c_{j}+c^{\dag}_{j}c_{i}\right).
\eea
Here $\epsilon_{i}$ is a random onsite potential and $t_{ij}$ is the hopping strength. For simplicity we take $\epsilon_i$ to be uniformly distributed in $\left[-W,W\right]$ with density $\rho(\epsilon)=\frac{1}{2W}\Theta(W-|\epsilon|)$ and choose energy units such that $W=1$. The operators $c_{i}(c^{\dag}_{i})$ create or annihilate a fermion at the lattice site $i$. 

The phenomenology of this canonical model is well established, due to extensive analytical and numerical studies.
In 3d, at weak disorder, most eigenstates are delocalized and form a continuum that touches the localized states in the tails of the band at the so-called mobility edges. Upon increase of the disorder the mobility edges move towards the bulk of the spectrum. The states in the middle of the band, where the density of states is highest, are usually the last to localize.
On a 3d cubic lattice, this happens when the hopping becomes weaker than $t_c \approx 
0.12$.~\cite{3DAnderson} The single particle wavefunctions at the mobility edge are neither fully space-filling nor fully localized, but exhibit interesting multifractal properties.~\cite{MirlinEvers}

Given the canonical model (\ref{Anderson}) of localization of free fermions, it is interesting to study what changes if we replace fermionic with bosonic operators $c_{i}(c^{\dag}_{i})\to b_{i}(b^{\dag}_{i})$. However, as mentioned above, non-interacting disordered bosons exhibit pathological behavior, since they simply condense into the lowest lying single particle wavefunction, which is generically a strongly localized state at the extreme of the Lifshitz tail of the density of states. To remedy this pathology, we must include interactions, a
particularly interesting case being {\em hard-core} bosons which locally repel each other infinitely strongly.
This model has local constraints like spinless fermions, which obey the Pauli exclusion principle.
In both cases at most one particle can occupy a given lattice site. The two models differ, however, 
due to the exchange statistics. In the case of hard core bosons, the local repulsion renders the system genuinely interacting, 
while `hard core' fermions can of course be understood entirely by solving the single particle problem at all energies. 

The difference in the quantum statistics is ultimately responsible for the fact that superfluids of bosons survive weak disorder in spatial dimensions $d=2$, whereas repulsive fermions are generically prone to localize and form insulators at low temperature.
How precisely a disordered Bose glass turns into a delocalized superfluid, especially in low dimensions, is not understood in full detail, even though there has been recent progress on the experimental front, as well as in numerical simulations.~\cite{Prokofev, 3ddisoerderedhardcorebosons, Roscilde}
Questions regarding the localization of excitations in the Bose glass, the existence of bosonic mobility edges, 
or a finite temperature \textquotedblleft many-body delocalization\textquotedblright~in bosons are being debated, too, and serve as a motivation for the present analysis.

\subsection{Realization of hard core bosons}
Physical realizations of hard core bosons arise naturally in several contexts: Apart from the obvious example of strongly 
repulsive cold bosonic atoms, hard core bosons emerge in correlated materials with a strong local negative $U$ attraction 
where all electron sites are either empty or host two electrons of opposite spin. Naturally, such singlets form hard core bosons.
A minimal description in the presence of disorder is given by the Hamiltonian
\bea
\label{hardcorebosons}
H_{\rm hcb}=-\sum_{i}\epsilon_{i}n_{i}- \sum_{\langle i,j \rangle}t_{ij} \left( b^{\dag}_{i}b_{j}+b^{\dag}_{j}b_{i}\right),
\eea
where interactions are retained only in the form of a local hard core constraint.
Such a Hamiltonian was also obtained via an approximate description of strongly disordered superconductors by Ma and Lee,~\cite{MaLee}
who generalized the BCS wavefunction to be built from doubly occupied or empty single particle wavefunctions.
Each such orbital thus forms an Anderson pseudospin, that is, a hard core boson. The orbitals will be localized in 
space if the disorder is strong. The Ma-Lee model allows one to describe approximately the 
superfluid-to-insulator transition in strongly disordered systems with predominant attractive interactions.
Recently this approach has been extended to take into account the fractality of paired single particle states close to 
an Anderson transition,~\cite{FeigelmanIoffeKravtsov, BurmistrovMirlin} which translates into unusual statistics of the pair 
hopping elements $t_{ij}$ occurring in (\ref{hardcorebosons}).

In the past, the thermodynamics of the Hamiltonian (\ref{hardcorebosons}) was studied extensively with quantum 
Monte Carlo techniques~\cite{SItransition,SItransition2d, 1dBoseglass, Zhang} as a model for the disorder driven 
superfluid-insulator transition. Recently the model was revisited from the perspective of localization of excitations 
in the insulating regime.~\cite{IoffeMezard,FeigelmanIoffeMezard,Markus} Interestingly, clear signs of the quantum statistics,
 i.e. differences between fermions and hard core bosons, appear already deep in the insulating regime:~\cite{positivemagnetoresistance,BapstMueller} 
In $d>1$ one finds that low energy excitations of hard core bosons delocalize more readily than fermions when subjected to 
the same disorder potential. A more important difference is the fact that the wavefunctions of localized excitations react 
in opposite ways to a magnetic field: While fermionic excitations tend to become more delocalized due to the suppression 
of negative interference of alternative tunneling paths, bosonic
wavefunctions tend to contract under a magnetic field. This leads to strong, opposite magnetoresistance in the 
low temperature transport of such insulators.~\cite{ZhaoSpivak,SpivakShklovskii,positivemagnetoresistance,Gangopadhyay2012}

\subsection{Random field magnets}
The Hamiltonian of disordered bosons (Eq.~(\ref{hardcorebosons})) is equivalent to a XY ferromagnet of $s=1/2$ spins in a random transverse field. This is easily seen using the isomorphism  $b_{i}=\sigma^{-}_{i}=\frac{1}{2}(\sigma^{x}_{i}-i\sigma^{y}_{i})$, $b^{\dag}_{i}=\sigma^{+}_{i}=\frac{1}{2}(\sigma^{x}_{i}+i\sigma^{y}_{i})$, $n_{i}=(\sigma^{z}_{i}+1)/2$:

\bea
\label{disorderedXY}
H_{\rm XY} &=&-\sum_{i}\epsilon_{i}\sigma^{z}_{i}-J \sum_{\langle i,j \rangle}\left(\sigma^{x}_{i}\sigma^{x}_{j}+\sigma^{y}_{i}\sigma^{y}_{j}\right),\nn\\
&=& 2H_{\rm hcb}+{\rm const.}\,,
\eea
where we took the hopping or exchange couplings to be uniform, $t_{ij}=J$.
In Eq.~(\ref{disorderedXY}) the operators $\sigma^{x,y,z}_{i}$ are Pauli matrices. At zero temperature, this model exhibits a quantum disordered Bose glass (paramagnetic) phase for $t ,J\ll W$. In dimensions $d>1$, a superfluid (ferromagnetic) phase is expected for $t,J\gg W$, while strictly one-dimensional chains are known to be fully localized irrespective of the weakness of disorder.

We will contrast the model (\ref{disorderedXY}) with the closely related Ising model
\bea
\label{disorderedIsing}
H_{\rm Ising}=-\sum_{i}\epsilon_{i}\sigma^{z}_{i}- J\sum_{\langle i,j \rangle} \sigma^{x}_{i}\sigma^{x}_{j}.
\eea
Note that while the hardcore boson model conserves particle number and possesses the related continuous $U(1)$ symmetry,
the model (\ref{disorderedIsing}) only conserves the parity of the total $z$-component of the spin $\sum_i \sigma^z_i$ and possesses the global discrete Ising symmetry $\sigma^x\to -\sigma^x$. 
For brevity we shall refer to the above models as the XY and Ising models, respectively. 

It is one of the main goals of this paper to analyze the effect of the different symmetries on the properties of excitations in the localized phases and on the approach to the ordered phase. %, even though at first glance one might expect that the quantum fluctuations introduced by the exchange interactions $\sim J$ have a very similar effect.

\section{Review of previous results}
\label{sec:review}
In order to situate the present study in the context of the existing literature on disordered bosonic systems, 
we briefly review  previous analytical approaches and their key results, before giving a short overview of the results of this paper.

\subsection{1-dimensional systems}
Many previous studies have focused on the properties of ground states and the quantum phase transition between insulating 
and superfluid phases of 1d spin chains and bosons at $T=0$. 
Strongly interacting Bose gases in weak disorder in 1d are amenable to a Luttinger liquid description.~\cite{GiamarchiSchulz, GiamarchiBook} 
The superfluid-insulator quantum phase transition occurs at a universal value $K=3/2$ of the Luttinger parameter,~\cite{Risti} 
corresponding to an unstable fixed point of a Kosterlitz-Thouless-type renormalization group flow. 
In the opposite limit of strong disorder and moderate interactions, complementary approaches suggest 
that the quantum phase transition is still present,~\cite{SanchezPalencia, Nattermann, Altman} but occurs at a different, 
non-universal value~\cite{Altmanstrongdisorder} of $K$, and the transition may thus belong to a different universality class. 
While this was supported by the recent numerical study,~\cite{HrahshehVojta} the later work~\cite{Pollet} argued that in strong disorder the true critical behavior emerges only at  exponentially large scales, predicting the universal value of $K_c=3/2$ to hold even in that limit. 
An attempt to extend the analysis of the critical point from $d=1$ to higher dimensions via an $\epsilon$-expansion was made in Ref.~\onlinecite{Herbut}.

At {\em finite temperatures}, the weakly interacting, disordered Bose glass has been argued to undergo a localization transition 
between a fluid phase with finite d.c. transport at high temperature and an ideal insulator 
with no conduction at low $T$.~\cite{finitetemperature} Hereby the transition temperature plays the role of an extensive 
mobility edge separating localized and delocalized states of the system. 
In Ref.~\onlinecite{finitetemperature} it has been conjectured that in $d=1$ the transition temperature is non-zero 
everywhere in the Bose insulator, with a $T_c$ that tends to zero at the superfluid transition.  

Closely related disordered spin chains have been studied in rather great detail, see e.g., Ref.~\onlinecite {DotyFisher}.
An important, canonical example is the random transverse field Ising chain~(\ref{disorderedIsing}). 
Its disorder-driven transition from paramagnet to ferromagnet and the associated critical behavior can be described by the asymptotically exact real space renormalization group (RSRG),~\cite{Fisher} which flows towards infinitely strong randomness at the fixed point. An important hallmark of this type of fixed points is activated dynamical scaling (characteristic frequencies scaling exponentially with the relevant length scales) which formally corresponds to an infinite dynamical critical exponent.~\cite{VojtaReview} 
An important insight from these studies is the fact that at the critical point {\em average} and {\em typical} observables behave very differently, since averages are dominated by exponentially rare events. The difference between average and typical observables persists also
in the off-critical regions in the form of gapless Griffith phases.
 
\subsection{Higher dimensions $d>1$}
In strictly one-dimensional chains the disordered XY model Eq.~(\ref{disorderedXY}) is always in a localized phase, as a Jordan-Wigner transformation maps it to free, disordered fermions.  
In higher dimensions XY models can however develop true long-range order, as was shown by numerical studies on disordered hard core bosons.~\cite{SItransition,SItransition2d,3ddisoerderedhardcorebosons}
The transport properties of the Bose glass phase remains however  a non-trivial and controversial problem, especially close to the superfluid transition, where the question as to the existence of many body mobility edges and their relevance for purely bosonic conductivity arises.~\cite{Markus,finitetemperature,IoffeMezard,FeigelmanIoffeMezard, AlvarezLaflorencie}

For Ising models, in higher dimensions the RSRG procedure can still be employed. However, it can only be carried out numerically as the effective lattices generated upon decimation become more and more complex.
Numerical studies in dimensions $d=2,3,4$, as well as on regular graphs, have found that transverse field Ising models
are still governed by infinite randomness fixed points, which justify the RSRG procedure {\em a posteriori}, at least close to criticality.~\cite{Motrunich, KovacsIgloi2d, KovacsIgloi3d} In 2d, the conclusions about the critical behavior have been verified by quantum Monte Carlo (QMC) simulations.~\cite{Pich, Rieger} The RSRG of Ref.~\onlinecite{Motrunich} suggests that the phase transition occurs as a kind of a percolation-aggregation phenomenon of spin clusters, which form a system-spanning cluster of low fractal dimension $d_f<d$ at criticality ($d_f \approx 1$ in 2d).

Deep in the disordered phase of Ising models low energy excitations have been studied within leading order perturbation theory in the exchange.~\cite{IoffeMezard,FeigelmanIoffeMezard, MonthusGarel,positivemagnetoresistance} This approach will be revisited and put on firm ground in this paper, extending the analysis  to finite excitation energies. 
In dimensions $d>1$ the lowest order perturbation theory maps off-diagonal susceptibilities in the transverse field Ising model to directed polymer problems, which have been studied extensively. Griffith phenomena are obtained naturally within such an approach as well. If the approximation is pushed all the way to the quantum phase transition (disregarding that perturbation theory becomes uncontrolled in low dimensions) unconventional activated scaling is found, similarly as expected from infinite randomness.~\cite{MonthusGarel} 
For XY models, the exact mapping between the leading order perturbation theory for boson Green's functions and directed polymers has been used 
to analyze the magnetoresistance in the insulating phase of charged hard core bosons.~\cite{Gangopadhyay2012}

Regarding the long range ordered side of these systems, excitations of the strongly inhomogeneous ordered Ising phase were analyzed in Refs.~\onlinecite{DimitrovaMezard,extendedcavity} by employing a "cavity" mean 
field approximation, to obtain spatial order parameter distributions. However, the status of this approximation remains unclear.
On the superfluid side of strongly disordered bosons (or XY spins), a variety of approaches have been used to address the properties 
of Bogoliubov excitations.~\cite{ChalkerGurarie,CordMueller, AlvarezLaflorencie}

\subsection{The role of order parameter symmetry}
RSRG studies in higher dimensions have pointed towards an important difference between Ising models and models with continuous spin symmetry: While the former were found to exhibit infinite randomness fixed points with activated scaling, models with continuous symmetry are found to have more conventional power law scaling at their quantum critical points.~\cite{Motrunich, Melin, VojtaReview} This was confirmed by Quantum Monte Carlo simulations for Heisenberg models.~\cite{Laflorencie} The recent work by Feigelman et al.~\cite{FeigelmanIoffeMezard} suggested that in high dimensions, or at least on the Bethe lattice, the difference between the models studied here is inessential. However, we will show that a closer analysis does distinguish the two symmetries, and reveals their rather different critical properties.  

\subsection{Summary of new results}

In this paper, we focus on the models~(\ref{disorderedXY},\ref{disorderedIsing}), defined on general lattices. 
We first consider them deep in their insulating (disordered) phase, 
where we can safely work at low orders of perturbation theory in small hopping or exchange. 
We study the localization properties of excitations, 
by analyzing the lifetime of excitations due to an infinitesimal coupling to a bath at the boundaries of the sample, 
which allows us to characterize the exponential localization of bulk excitations. 
Comparing this perturbative approach to exact results for one-dimensional Ising and XY chains, 
we obtain insight on the relevance of the most important subleading corrections, and use them subsequently to analyze higher dimensional lattices, 
in particular highly connected Bethe lattices.  

As shown in Ref.~\onlinecite{positivemagnetoresistance} in the insulator, excitations at low energies extend the farther in space the lower their energy, both in Ising and XY models, 
provided the disorder is uniformly distributed. 
This phenomenology is in qualitative agreement with RSRG approaches (in regimes where those apply), 
in that the real space decimation constructs lower and lower energy excitations with increasing spatial extent. 
This distinguishes the bosonic models from non-interacting fermions, 
which are rather insensitive to the excitation energy. Furthermore,
for Ising chains we find a non-analytic behavior, $\xi^{-1}(\omega)-\xi^{-1}(0)\sim |\omega|^\alpha$
of the localization length at small excitation energies, where $\alpha>0$ vanishes upon approaching the phase transition. 
This exponent remains small in a substantial interval around the critical point before it eventually saturates to $\alpha=2$, 
corresponding to analytical behavior. We predict similar behavior also for higher dimensions. 
This is closely related to the Griffith phenomena found by the RSRG approach for Ising models.
 
In order to approach the quantum phase transition to the long range ordered phase,
we apply our formalism to Cayley trees of large connectivity, where a perturbative approach is controlled up to a parametrically 
small vicinity of the transition. Moreover, we argue that the phenomenology of the transition is correctly captured by summing 
the most relevant subleading contributions in the  perturbative approach. In this limit we find that independently of 
the symmetry of the model, the localization length at non-extensive low energy excitations remains a local maximum at $\omega=0$, 
all the way to the phase transition. In the Ising model, we argue that this statement holds in a {\em finite} range of frequencies up the transition, while in the XY model we control only a regime of frequencies which shrinks to zero as the transition is approached. 
For the Ising model our findings suggest that the ordering transition occurs by a delocalization {\em initiated at $\omega=0$}, 
i.e., without the preceding delocalization of excitations at small positive energies. In both considered models, in the presence of uniform disorder 
potential, we did not find evidence for a mobility edge of bosonic excitations at finite energies on the insulating side. However, close to criticality, we cannot exclude the existence of such a mobility edge at higher intensive energies, 
since our perturbative approach to low orders  cannot fully describe the propagation of larger lumps of energy. 

However, a mobility edge is found rather trivially at higher energies (at least in $d>2$) for non-uniform disorder, if the density of disorder energies increases away from the chemical potential.
Our results for the Bethe lattice indicate that the percolating structure on which the emerging 
long range order establishes is significantly more sparse for random transverse field Ising models than for systems with XY symmetry.

\section{Locator expansion}
\label{s:decay}
\subsection{Decay rate of local excitations}
In this Section we study the decay of local excitations (spin flips) sufficiently deep within the disordered phases -- the insulating Bose glass or the paramagnet --
of the models (\ref{disorderedXY})  and (\ref{disorderedIsing}), respectively. 
The transverse quantum fluctuations due to the  exchange $J$ (or hopping $t$) allow spin flips to propagate over some distance. 
However, within a localized regime, they die off exponentially at large distance. 
Following, in spirit, Anderson's approach to single particle localization, 
we characterize localization by the  decay rate $\Gamma$ of a local excitation, 
as induced by the coupling to a bath at the distant boundaries of the sample. 
In the localized phase $\Gamma$ is exponentially small in the linear size of the system. 
A good measure for the localization radius $\xi$ of such excitations is thus given by the decrease of $\log \Gamma$ with 
the distance $R$ to the boundary, which generally behaves as $\log \Gamma \approx - 2R/\xi$.
For delocalization, and thus energy diffusion, to occur in a many body context, 
typical decay rates must remain finite in the thermodynamic limit, as the system size tends to infinity, 
while the coupling to the bath is kept infinitesimal.

We study the models (\ref{disorderedXY}) and (\ref{disorderedIsing}) on a general lattice $\Lambda$. 
We assume the system to be coupled infinitesimally to a zero temperature bath via the spins 
on a \textquotedblleft boundary set\textquotedblright~ $\partial \Lambda$ of the lattice, 
which becomes infinite in the thermodynamic limit, too. 
Later on, to simplify the discussion, we will choose this subset to be the spatial boundary of the finite lattice $\Lambda$.~\cite{fn1} 
All boundary sites $l\in \partial \Lambda$ are assumed to be coupled to independent, identical baths, described by a continuum of
non-interacting harmonic oscillator modes $b_{\alpha,l}$ of energy $\epsilon_\alpha$ and coupling strength $\lambda_\alpha$:
\bea
H_{\textrm{b}} &=& \sum_{l\in \partial \Lambda}\sum_{\alpha}\epsilon_{\alpha}b^{\dag}_{\alpha,l}b_{\alpha,l}.
\eea
 Such a bath is characterized by its spectral function
\bea
J_b(\epsilon)=\sum_{\alpha}\lambda^{2}_{\alpha}\delta(\epsilon-\epsilon_{\alpha}).
\eea

For both the XY and Ising models we consider the following system-bath couplings:
\bea
\label{randomspinmodelswithbath}
H &=&H_0 +H_{\textrm{s,b}}+H_{\textrm{b}},\nn\\
H_{\textrm{s,b}} &=&
\sum_{l\in\partial \Lambda} \sigma^{x}_{l}\sum_{\alpha}\lambda_{\alpha}\left(b^{\dag}_{\alpha,l}+
b_{\alpha,l} \right),
\eea
where $H_0 = H_{\rm XY,Ising}$ is the uncoupled spin Hamiltonian. Obviously, 
the details of the coupling to the bath are irrelevant for the determination of localization radii $\xi$ of localized excitations, or to determine the presence of delocalization.

In the limit $J \ll1 $, the ground state is well approximated by the product state
\bea
|\textrm{GS}\rangle \approx \otimes_{i\in \Lambda} \left|\sigma_i^z= {\rm sign}(\epsilon_i)\right\rangle\,.
\eea
Let us now characterize the temporal decay of a local excitation close to the site $0\in \Lambda$ in the bulk of the lattice. As a canonic example we will study the spin flip excitation $\sigma^{x}_{0}|\textrm{GS}\rangle$.  For $J=0$ this creates the excited state
\bea
 |E^{0}\rangle =
 \otimes_{i\in \Lambda} \left | \sigma_i^z= (1-2\delta_{0i}){\rm sign}(\epsilon_i) \right\rangle.
\eea

At finite $J$ we denote by the same ket $|E^0\rangle$ the eigenstate, which evolves adiabatically 
from the excited state (at $J=0$) and thus has largest overlap with the local spin flip excitation at small $J$. 
Our aim is to determine the lifetime of that eigenstate in the limit of large system size. 
In large but finite systems, the lifetime is finite since the coupling to the bath induces decays to lower energy states, and in particular back to the ground state.
As an explicit calculation below will confirm the lifetime can be evaluated simply by applying Fermi's Golden rule.

One should naturally ask whether the lifetime of spin flip excitations (which carry Ising or $U(1)$ charge) should be characteristic for the lifetime of other excited states that are created by local operators. Among the excitations that transform the same way under Ising or XY symmetry operations, we expect that deep in the disordered phase, the localization length (as defined via the exponentially small inverse lifetime) is the same function of energy as that of single spin flips. This is because the propagation to long distances proceeds furthest by making the minimal use  of exchange couplings. At energies below the bandwidth this is always achieved by the shortest chains of exchange bonds between the location of excitation and the point of observation, which will be analyzed in detail below. For excitations with different symmetry, e.g. neutral ones, such as a pair of opposite spin flips, the localization is stronger in the regime of small $J$, since the matrix element to transport such an excitation by a
distance $R$ decays as $\sim J^{2R}$, as compared to the amplitude $\sim J^R$ for single spin flips. However, we do not know whether this property remains generally true all the way to the ordering transition, where the expansion in $J$ starts to diverge. There an estimate of the relative importance of various propagation channels can be very difficult, and depends on the details of the considered model. Throughout this paper we thus stay within regimes where the perturbative expansion in the small exchange $J$ is controlled.

In order to make the above notions formally precise, we define the retarded spin correlator
\bea
\label{Greenfunctiondefinition}
G_{l,0}(t)\equiv-i\Theta(t)\, _{b}\langle \textrm{GS}|[\sigma^{x}_{l}(t),\sigma^{x}_{0}]|\textrm{GS}\rangle_{b},
\eea
where $|\textrm{GS}\rangle_{b}=|\textrm{GS}\rangle\otimes|\textrm{bath} (T=0)\rangle$
denotes the ground state of the uncoupled system, $|\textrm{{GS}}\rangle$ being the ground 
state of $H_{0}$, $H_{0}|\textrm{{GS}}\rangle=E_{\textrm{GS}}|\textrm{GS}\rangle$, 
and $|\textrm{bath } (T=0)\rangle$ being the ground state of the bath.
$A(t)= e^{-iHt}Ae^{iHt}$ denote Heisenberg operators. In the following, we will analyze in particular local correlators, such as
$G_{0,0}(t)$. It will be convenient to study these correlators in the frequency domain
\bea
\label{FourierTrafo}
G_{l,0}(\omega)=\int^{\infty}_{-\infty} dt \textrm{e}^{i(\omega+i\eta) t}G_{l,0}(t),
\eea
with $\eta\rightarrow 0^{+}$. Introducing ${\cal U}(t)=\textrm{e}^{iH_{0}t}\textrm{e}^{-iHt}$, we can write
\bea
\label{localGreenfunction}
&&G_{0,0}(t)=\\
&&\quad\quad -i\Theta(t)\,_{b}\langle \textrm{GS}|[{\cal U}^{\dag}(t)\textrm{e}^{iH_{0}t}\sigma^{x}_{0}\textrm{e}^{-iH_{0}t}{\cal U}(t),\sigma^{x}_{0}]|\textrm{GS}\rangle_{b}. \nn
\eea
We evaluate (\ref{localGreenfunction}) perturbatively in the coupling to the bath, expanding ${\cal U}(t)$ in $\lambda_\alpha$. To second order one finds
\bea
{\cal U}(t)&\simeq& 1-i\int^{t}_{0}dt_{1}H_{\textrm{s,b}}(t_{1})\\
&&-\int^{t}_{0}dt_{1}\int^{t_{1}}_{0}dt_{2} H_{\textrm{s,b}}(t_{1})H_{\textrm{s,b}}(t_{2})+O(\lambda_\alpha^3)\nn,
\eea
where $H_{\textrm{s,b}}(t)=\textrm{e}^{iH_{0}t}H_{\textrm{s,b}}\textrm{e}^{-iH_{0}t}$.
Inserting into $G_{0,0}(\omega)$ we obtain the expansion
\bea
G_{0,0}(\omega)=G^{(0)}_{0,0}(\omega)+G^{(2)}_{0,0}(\omega)+o(\lambda^{2}_{\alpha}),
\eea
whereby the linear term in $\lambda_\alpha$ vanishes due to conservation of the parity of the total spin projection, $\sum_{i}\sigma^{z}_{i}$.
 The leading term is
\bea
G^{(0)}_{0,0}(\omega)=\sum_{n}\left(\dfrac{|\langle \textrm{GS}|\sigma^{x}_{0}|E_{n}\rangle|^{2}}{\omega+E_{\textrm{GS}}-E_{n}+i\eta}-\right. \nn\\
\left.\dfrac{|\langle E_{n}|\sigma^{x}_{0}|\textrm{GS}\rangle|^{2}}{\omega+E_{n}-E_{\textrm{GS}}+i\eta}\right),
\eea
where $n$ runs over all eigenstates of $H_0$, which are 
labeled by their energy $E_{n}$.

The second order term $G^{(2)}_{0,0}(\omega)$ has a relatively complicated structure for arbitrary $\omega$.
However, we are particularly interested in understanding the lifetime of excitations, 
i.e., the imaginary part of the poles that appear in the leading term $G^{(2)}_{0,0}(\omega)$. 
Therefore we focus on $\omega\approx E_{n}-E_{\textrm{GS}}$, 
and extract only the most singular term in the imaginary part of $G^{(2)}_{0,0}(\omega \rightarrow E_{n}-E_{\textrm{GS}})$, 
which evaluates to:
\bea
&& \textrm{Im}\,G^{(2)}_{0,0}(\omega\rightarrow E_n-E_{\rm GS}) =
- \pi \dfrac{|\langle \textrm{GS}|\sigma^{x}_{0}|E_{n}\rangle|^{2}}{(\omega+E_{\textrm{GS}}-E_{n})^{2}}\nn\\
&&\quad  \times \sum_{l\in\partial \Lambda} \sum_{E_m<E_n} J_b(E_n-E_{m}) |\langle E_m|\sigma^{x}_{l}|E_{n}\rangle|^{2}\nn\\
&&\quad  \times \left[1+O(\omega+E_{\textrm{GS}}-E_{n})\right].
\eea
As the bath couplings and the spectral functions $J_b(\omega)$ are assumed to be very small, 
we can account for this imaginary part as a shift of the poles into the complex plane:
\bea
\label{G00}
G_{0,0}(\omega)\approx \sum_{n}\dfrac{|\langle \textrm{GS}|\sigma^{x}_{0}|E_{n}\rangle|^{2}}{\omega-(E_{n}-E_{\textrm{GS}}-i \Gamma_n/2)},
\eea
where, to quadratic order in the bath coupling,
\bea
\Gamma_n &=&  2\pi \sum_{l\in\partial \Lambda} J_b(E_n-E_{\rm GS}) |\langle \textrm{GS}|\sigma^{x}_{l}|E_{n}\rangle|^{2} \nn\\
&& +2\pi \sum_{l\in\partial\Lambda} \sum_{E_{\rm GS}<E_m<E_n} J_b(E_n-E_{m}) |\langle E_m|\sigma^{x}_{l}|E_{n}\rangle|^{2} \nn
\eea
is the decay rate of the excited state $n$ under emission of a bath mode. 
This is easily recognized as the inverse lifetime expected from Fermi's Golden rule.
We have dropped the real parts of the self-energies which shift the poles by small amounts proportional to the coupling to the bath.

Note that the rate $\Gamma_n$ includes the decay to the ground state as well as to other excited states of lower energy. 
However, we will merely focus on the contribution from the decay to the ground state,
\bea
\label{GammaGS}
\Gamma_n^{({\rm GS})} &=&  2\pi J_b(E_n-E_{\rm GS}) \sum_{l\in\partial \Lambda}  |\langle \textrm{GS}|\sigma^{x}_{l}|E_{n}\rangle|^{2} .
\eea
This indeed suffices for our purposes, for two reasons:
On one hand, if we are interested in the rate at which energy escapes the system, 
we should not consider decays to other excited states, since those retain part of the excitation energy within the system. 
On the other hand, the relevant contributions to the full inverse lifetime due to decays into excited states are 
comparable in magnitude to the contribution from the decay to the ground state. Therefore the latter furnishes enough information to determine the localization radius  of the excitations in a deeply insulating regime.

As we explained before, we are interested in the excited state $|E_n\rangle =|E^0\rangle$ which dominates the many body wavefunction after a spin flip at site $0$.  
To compute its lifetime, according to (\ref{GammaGS}), we need to evaluate the matrix element
\bea
\langle \textrm{GS}|\sigma^{x}_{l}|E^{0}\rangle\approx \langle \textrm{GS}|\sigma^{x}_{l}|E^{0}\rangle
\langle  E^{0}|\sigma^{x}_{0}| \textrm{GS}\rangle\equiv A_{l0},
\eea
since $\langle  E^{0}|\sigma^{x}_{0}| \textrm{GS}\rangle = 1-O(J^2)$. 
Note the asymmetry of $l$ and $0$ in the definition of this amplitude: $A_{l0}\neq A_{0l}$. 
The right index ($0$) denotes the site where the excitation is created, 
while the left index $l$ is the site at which the excitation is probed.

From a Lehmann representation of the retarded Green's function (\ref{Greenfunctiondefinition}) 
it becomes clear that $A_{l0}$ is simply one of its residues:
\bea
&&G_{l,0}(\omega)
%\equiv \int^{\infty}_{-\infty} G_{l,0}(t) e^{i(\omega+i\eta)t}dt \nn\\
= \int^{\infty}_{-\infty} -i\Theta(t)\langle \textrm{GS}|\left[\sigma^{x}_{l}(t), \sigma^{x}_{0}\right]|\textrm{GS}\rangle e^{i(\omega+i\eta)}dt \nn\\
&&\quad=\sum_{n} \langle E_{n}|\sigma^{x}_{0}|\textrm{GS}\rangle\langle\textrm{GS}|\sigma^{x}_{l}|E_{n}\rangle \times \nn\\
&&\quad \times \left[\frac{1}{\omega-(E_{n}-E_{\textrm{GS}}-i\eta)} 
%&&\left.\quad
-\frac{1}{\omega+(E_{n}-E_{\textrm{GS}}+i\eta)}\right].\nn
\eea
Indeed, the residue of the pole at $\omega = E^0-E_{\rm GS}$ is the desired matrix element,
\bea
A_{l0}= \lim_{\omega \to E^0-E_{\rm GS}} [\omega - (E^0-E_{\rm GS})] G_{l,0}(\omega).
\eea
This observation can be used to determine $A_{l0}$ in perturbation theory in $J$ 
in an efficient manner, by solving recursively the equation of motion for $G_{l,0}$ in a locator expansion.
As we will discuss further below, the matrix elements $A_{l0}$ are needed 
not only to calculate the decay rate to the ground state, 
but also to determine the onset of long range order.

The above derivation is easily adapted for the XY model, cf. Sec.~\ref{ss:XY}.

\subsection {Equations of motion - Ising model}
Let us now evaluate the Green's function $G_{l,0}(t)$ to the leading order in the exchange $J$. 
We first split $G_{l,0}(t)$ into two parts
\bea
\label{Gsplit}
G_{l,0}(t)&=&G^{+}_{l,0}(t)+G^{-}_{l,0}(t),\\
G^{\pm}_{l,0}(t)&=& -i\Theta(t)\langle \textrm{GS}|\left[\sigma^{\pm}_{l}(t),\sigma^{x}_{0}\right]|\textrm{GS}\rangle,\nn
\eea
which satisfy simpler equations of motion,
\bea
\label{EqofM}
i\frac{dG^{\pm}_{l,0}(t)}{dt}=\delta(t) \langle \left[\sigma^{\pm}_{l}(t), \sigma^{x}_{0}\right] \rangle-i\Theta(t)\langle\left[i\dot{\sigma}^{\pm}_{l}(t), \sigma^{x}_{0}\right] \rangle.
\eea

The spin flip operators $\sigma^{\pm}_{l}(t)$ satisfy Heisenberg equations. For the Ising model, they read
\bea
i\dot{\sigma}^{\pm}_{l}(t) = \left[\sigma^{\pm}_{l}, H_{0}\right]
=\pm 2\epsilon_{l}\sigma^{\pm}_{l}(t)\mp J \sigma^{z}_{l}(t)\sum_{j\in\partial l}\sigma^{x}_{j}(t).\,
\eea
The sum is over the set $\partial l$ of nearest-neighbors of site $l$. To leading order in $J$, we can restrict ourselves to the neighbors $j$ whose distance to site $0$ is smaller than that of site $l$. Other terms lead to contributions of higher order in $J$. Furthermore, when evaluating the expectation value in the last term of Eq.~(\ref{EqofM}), we can decouple the average over $\sigma^{z}_{l}(t)$ from the other operators,
\bea
\label{decoupling}
\langle\sigma^{z}_{l}(t) \sigma^x_j ... \rangle = \langle\sigma^{z}_{l}(t)\rangle \langle \sigma^x_j ... \rangle\,,
\eea
and use $\langle\sigma^{z}_{l}(t)\rangle=\textrm{sign}(\epsilon_{l})+O(J^{2})$.
Corrections to this approximation lead again to higher powers of $J$. 
They can be determined systematically by an extension of the present approach.~\cite{BapstMueller}

To the leading order in the exchange, the recursion relations for the Green's functions, after Fourier transform, become
\bea
(2\epsilon_{l}\mp \omega)G^{\pm}_{l,0}=J\, \textrm{sign}(\epsilon_{l})\sum_{j\in\partial l}G_{j,0}(\omega).
\eea
Solving for $G_{l,0}(\omega)$ from Eq.~(\ref{Gsplit}) we obtain the recursion relation
\bea
\label{recursionrelation}
G_{l,0}(\omega)=\sum_{j \in\partial l}J\, \textrm{sign}(\epsilon_{l})\frac{4\epsilon_{l}}{(2\epsilon_{l})^2-\omega^{2}} G_{j,0}(\omega),
\eea
which is exact to leading order in $J$.
Upon iterating the recursion until we reach the site $0$, we obtain the leading order of the Green's function as a sum over all shortest paths from $l$ to $0$ (of length $L={\rm dist}(l,0)$, the Hamming distance on the lattice between $l$ and $0$),
\bea
\label{matrixelement}
G_{l,0}(\omega)&=&G_{0,0}(\omega)\sum_{{\cal P}=\{j_{0}=0,..,j_{L}=l\}}\prod^{L={\rm dist}(l,0)}_{p=1}
\frac{4J |\epsilon_{j_{p}}|}{(2\epsilon_{j_{p}})^{2}-\omega^{2}} \nn\\
&&\quad\quad +o(J^L).
\eea
Notice that $G_{0,0}(\omega\rightarrow E^{0}-E_{\textrm{GS}})\approx\frac{1}{\omega+E_{\textrm{GS}}-E^{0}}$ to zeroth order in $J$. 
Therefore, the sought residue of the pole at $\omega=E^{0}-E_{\textrm{GS}}= 2|\epsilon_0|+O(J^2)$ in $G_{l,0}(\omega)$ is
\bea
\label{Isingmatrixelement}
A_{l0}&=& \left. \dfrac{G_{l,0}(\omega)}{G_{0,0}(\omega)}\right|_{\omega=2|\epsilon_0|}\\
&=&\sum_{{\cal P}=\{j_{0}=0,..,j_{L}=l\}}\prod^{L}_{p=1}
\frac{J|\epsilon_{j_{p}}|}{\epsilon_{j_{p}}^{2}-\epsilon_0^{2}}+o(J^L),\nn
\eea
to the leading order in $J$. An alternative derivation of this result by standard perturbation theory is given in the Appendix for the special, but non-trivial case of a three site Ising chain. 

\subsection {Equations of motion - XY model}
\label{ss:XY}
It is straightforward to repeat the same steps for the XY model. 
Without loss of generality, we suppose that the flipped spin sits on a site $0$ 
with $\epsilon_{0}\geq 0$ and thus essentially points up in the ground state.
We aim at the matrix element of the operator $\sigma^x_l$ between the ground state 
and the excited eigenstate $|E^{0}\rangle=  \sigma_0^-|{\rm GS}\rangle$ (up to corrections of order $O(J^2)$),
\bea
\langle \textrm{GS}|\sigma^{x}_{l}|E^{0}\rangle\approx \langle \textrm{GS}|\sigma^{+}_{l}|E^{0}\rangle
\langle  E^{0}|\sigma^{-}_{0}| \textrm{GS}\rangle\equiv A_{l0}.
\eea
We thus define the relevant Green's function as
\bea
G^{\rm XY}_{l,0}(t)\equiv -i\Theta(t)\langle \textrm{GS}|[\sigma^{+}_{l}(t),\sigma^{-}_{0}]|\textrm{GS}\rangle.
\eea
Employing the Lehmann representation and solving recursively the equations of motion in powers of $J$, 
allows us to extract the matrix element of interest
\bea
\label{XYmatrixelement}
A_{l0}&=&\left. \dfrac{G^{\rm XY}_{l,0}(\omega)}{G^{\rm XY}_{0,0}(\omega)}\right|_{\omega=2|\epsilon_0|}\nn\\
&=&\sum_{{\cal P}=\{j_{0}=0,..,j_{L}=l\}}\prod^{L}_{p=1}
\frac{J\textrm{sign}(\epsilon_{j_{p}})}{\epsilon_{j_{p}}-|\epsilon_0|}+o(J^L),\quad
\eea
to the leading order in powers of $J$.

Notice the difference between Eqs.~(\ref{Isingmatrixelement}) and~(\ref{XYmatrixelement}), 
which arises due to the different symmetries of the two models.
Indeed, in the Ising model, by a gauge transformation, one can always choose $\epsilon_{i}>0$, 
and therefore the physical correlators can only be functions of $|\epsilon_j|$, cf. Eq.~(\ref{Isingmatrixelement}). However, the same is not true for the XY model.
Within the leading order approximation the difference shows only at finite excitation energies $\omega\equiv 2|\epsilon_0|$, but disappears at low energies, 
$\epsilon_0\to 0$. 
This will be further discussed in Sec. IV  below. 

\subsection {Comparison with non-interacting particles (fermions)}
The result (\ref{XYmatrixelement}) was derived in Ref.~\onlinecite{positivemagnetoresistance} for hard core bosons. Using the correspondence $J\to t$, one obtains, up to subleading corrections,
\bea
 \langle \textrm{GS}|b^{\dagger}_{l}|E^{0}\rangle =  A_{l0},\nn
 \eea
where in this case we denote by
\bea
|E^{0}\rangle = b_0|{\rm GS}\rangle+O(t)\nn
\eea
the excited state with a boson removed from site $0$.
Note that Eq.~(\ref{XYmatrixelement}) has nearly the same form as the analogous sum for non-interacting fermions,~\cite{Anderson,SpivakShklovskii} 
which can be obtained from a recursive solution of the Schr\"odinger equation, 
or alternatively, using equations of motions as above for bosons. 
The fermion result differs only by the absence of the sign factors in the 
numerator ${\rm sign}(\epsilon)$. This is easy to understand physically:
Consider a loop formed by two shortest paths between two different sites $0$ and $l$. 
Taking the first path in forward direction and the second path backward, a ring exchange of particles is carried out. 
Indeed, in this process each particle is moving to the next negative energy site ahead of it on the loop. 
The corresponding amplitude for bosons and fermions should thus differ by an extra sign factor, 
if there is an odd number of particles  on the loop (in the ground state). 
This results precisely in the extra factor $\prod_{j\in {\rm loop}} \textrm{sign}(\epsilon_{j})$ which makes bosons distinct from fermions.

One should note that for non-interacting fermions the recursion relation for the Green's function is exact and does not require the decoupling (\ref{decoupling}) of the correlation function to obtain a closed recursion relation. Therefore the full Green's function can be expressed formally exactly as a sum over {\em all} paths, with amplitudes being products of the locators $t/(\epsilon_i-\omega)$, analogous to Eq.~(\ref{XYmatrixelement}), but without sign factors. In contrast, for hard core bosons this simple form is exact only for contributions from non-intersecting paths, where each link contributes a single factor of $J$. Loop corrections take a more involved form and require an extension of the equation of motion techniques used above.~\cite{BapstMueller}

\section{1-d case: localization in the Ising chain}
\label{s:1d}
In order to analyze certain features of higher dimensional cases, 
it is useful to review some properties of one-dimensional chains, 
which are exact solvable due to exact mappings to free fermions,~\cite{ShankarMurthy} and are even better understood with through the complementary approaches of bosonization ~\cite{GiamarchiSchulz} and the strong randomness renormalization group.~\cite{Fisher,Balents} 
The latter was first introduced for random transverse field Ising chains 
where it becomes asymptotically exact close to criticality.

On the other hand it is well-established that the random transverse field XY model, 
or hard core bosons in strictly one dimension does not exhibit a quantum phase transition, 
but only possesses the paramagnetic (insulating) phase. 
This can easily be seen from after a Jordan-Wigner transformation, 
which maps the hard core bosons to free fermions, which are always localized, 
even if the disorder potential is very weak. 
In bosonization language, hard core bosons in weak disorder are described 
by a Luttinger liquid with Luttinger parameter $K=1$. 
Only upon reducing the interactions strength, 
or compensating the hard core repulsion with an attractive interaction 
between bosons (equivalent to a $s^z_is^z_{i+1}$ interaction between spins) 
the value of $K$ can be increased above the critical value $K_c=3/2$, 
and a superfluid phase can emerge in sufficiently weak disorder.~\cite{GiamarchiSchulz,GiamarchiBook} 

In contrast to the XY spin chain without $s^zs^z$ coupling, 
the random transverse field Ising chain does undergo a 
para-to-ferromagnetic quantum phase transition, 
which is captured by an infinite randomness fixed point.~\cite{Fisher} 
We recall some of its properties that are relevant to our discussion for higher dimensions.
We consider the strictly one-dimensional model
\bea
\label{1dIsing}
H=-\sum_{j}\epsilon_{j}\sigma^{z}_{j}-J\sum_{j}\sigma^{x}_{j}\sigma^{x}_{j+1}.
\eea
As mentioned above, by a suitable gauge transformation, we can choose $\epsilon_{i}>0$.
Following the transformations and notations of Ref.~\onlinecite{3ddisoerderedhardcorebosons}, we introduce free fermions by the Jordan-Wigner transformation
\bea
c_{j}=\sigma^{-}_{j}e^{i\pi \sum_{k<j}\sigma^{+}_{k}\sigma^{-}_{k}},
\quad c^{\dag}_{j}=\sigma^{+}_{j}e^{-i\pi \sum_{k<j}\sigma^{+}_{k}\sigma^{-}_{k}}.
\eea
They satisfy the canonical anti-commutation relations
\bea
\{c^{\dag}_{j},c_{k}\}=\delta_{jk},\quad \{c^{\dag}_{j},c^{\dag}_{k}\}=\{c_{j},c_{k}\}=0.
\eea

In terms of fermionic degrees of freedom, the model (\ref{1dIsing}) can be written as

\bea
H=\sum_{jk}\left(\begin{array}{cc}
            c^{\dag}_{j}  & c_{j}
           \end{array}\right)
{\cal H}_{jk}\left(\begin{array}{c}
                                              c_{k} \\
                                              c^{\dag}_{k}
                                            \end{array}\right),
\eea
where
\bea
{\cal H}_{jk}=\frac{1}{2}\left(   \begin{array}{cc}
                                    {\cal D}_{jk}+ {\cal D}^{\dagger}_{jk}& {\cal D}_{jk}- {\cal D}^{\dagger}_{jk} \\
                                    {\cal D}^{\dagger}_{jk}- {\cal D}_{jk} & -{\cal D}_{jk}- {\cal D}^{\dagger}_{jk}
                                  \end{array}
    \right),
\eea
and ${\cal D}$ is the matrix defined as
\bea
\label{D}
{\cal D}_{jk}=-\epsilon_{j}\delta_{jk}-J\delta_{j,k-1}.
\eea
It is useful to perform a unitary transformation
\bea
(i\gamma^{1}_{j}, \gamma^{2}_{j})=(c^{\dag}_{j},c_{j}) U,
\eea

with
\bea
U=\frac{1}{\sqrt{2}}\left(\begin{array}{cc}
    1 & 1 \\
    -1 & 1
  \end{array}\right).
\eea
The operators $\gamma^{1(2)}_{j}$ are Majorana fermions corresponding to the imaginary(real) parts of $c^{\dag}_{j}$.
In terms of those the Hamiltonian takes the form
\bea
H=\sum_{jk}\left(\begin{array}{cc}
            i\gamma^{1}_{j}  & \gamma^{2}_{j}
           \end{array}\right)
\widetilde{{\cal H}}_{jk}\left(\begin{array}{c}
                                              -i\gamma^{1}_{k} \\
                                               \gamma^{2}_{k}
                                            \end{array}\right),
\eea
where
\bea
\widetilde{{\cal H}}=U^{-1}{\cal H}U=\left(\begin{array}{cc}
                       0 & {\cal D} \\
                       {\cal D}^{\dagger} & 0
                     \end{array}\right),
\eea
which makes explicit the chiral symmetry of the problem.
According to the classification of Altland and Zirnbauer,~\cite{Zirnbauer} the single particle Hamiltonian $\widetilde{{\cal H}}$ belongs to the chiral class BDI:~\cite{Balents,3ddisoerderedhardcorebosons} It is real, and there exists a matrix $\Sigma^{3}$, namely
\bea
\Sigma^{3}=\left(\begin{array}{cc}
    1 & 0 \\
    0 & -1
  \end{array}\right),
\eea
such that
\bea
\Sigma^{3}\widetilde{{\cal H}} \Sigma^{3} = -\widetilde{{\cal H}}˜.
\eea

\subsection{Transfer matrix approach}
For a chain of length $L$, there is a  unitary $2L\times 2L$ matrix $V$ which diagonalizes  $\widetilde{{\cal H}}$,
\bea
V \widetilde{{\cal H}} V^{-1} = {\rm diag}(\omega_1,...,\omega_{2L}),
\eea
and  its $n$'th column vector $(\psi^1_{n,i},\psi_{n,i}^2)$ satisfies the  Schr\"{o}dinger equation
\bea
\label{Schrodingerequation}
\left(\begin{array}{cc}
                       0 & {\cal D} \\
                       {\cal D}^{\dagger} & 0
                     \end{array}\right)\left(\begin{array}{cc}
                       \psi_n^{1} \\
                       \psi_n^{2}
                     \end{array}\right)=\omega_n \left(\begin{array}{cc}
                       \psi_n^{1} \\
                       \psi_n^{2}
                     \end{array}\right).
\eea
The corresponding operators
\bea
d_n = \sum_{i=1}^L \left[ \psi_{n,i}^1 (-i \gamma^1_i)+\psi^2_{n,i} \gamma^2_i\right]
\eea
and their conjugates satisfy canonical anti-commutation relations. They annihilate fermionic degrees of freedom of energy $\omega_n$, $[H,d_n] = -\omega_n d_n$.

In the lattice basis, Eq.~(\ref{Schrodingerequation}) takes the explicit form
\bea
\sum_{j}{\cal D}_{ij}\psi^{2}_{j}=\omega \psi^{1}_{i}, \\
\sum_{j}{\cal D}^{\dagger}_{ij}\psi^{1}_{j}=\omega \psi^{2}_{i},
\eea
where from here on we drop the mode index $n$.
Noting from (\ref{D}) that in these sums $j$ is restricted to the values $i-1,i$ or $i+1$, we find
\bea
{\cal D}_{ii+1}\psi^{2}_{i+1}=\omega\psi^{1}_{i}-{\cal D}_{ii}\psi^{2}_{i}-{\cal D}_{ii-1}\psi^{2}_{i-1},\\
{\cal D}^{\dagger}_{ii+1}\psi^{1}_{i+1}=\omega\psi^{2}_{i}-{\cal D}^{\dagger}_{ii}\psi^{1}_{i}-{\cal D}^{\dagger}_{ii-1}\psi^{1}_{i-1}.
\eea
Using Eq.~(\ref{D}) and the fact that ${\cal D}_{ii-1}={\cal D}^{\dagger}_{ii+1}=0$, this can be rewritten in the form of a recursive relation~\cite{ShankarMurthy,Fisher}

\bea
\left(\begin{array}{c}
  \psi^{1}_{i+1} \\
  \psi^{2}_{i+1}
\end{array}\right)=T_{i}(\omega)\left(\begin{array}{c}
                           \psi^{1}_{i} \\
                           \psi^{2}_{i}
                         \end{array}\right),
\eea
where the transfer-matrix $T_{i}(\omega)$ is given by
\bea
\label{transfermatrix}
T_{i}(\omega)&=&
\left(\begin{array}{cc}
-\frac{J}{\epsilon_{i+1}}\left(1-\frac{\omega^{2}}{J^{2}}\right)   & \frac{\omega \epsilon_{i}}{J \epsilon_{i+1}} \\
  -\frac{\omega}{J} & -\frac{\epsilon_{i}}{J}
\end{array}\right).
\eea

\subsection{Localization length}
Except at the critical point all the mode functions $\psi_i$ are exponentially localized. The typical localization length of these fermions, $\xi_{{\rm typ},f}$ can be extracted from the full transfer matrix ${\cal M(\omega)}=T_{L}(\omega)...T_{2}(\omega)T_{1}(\omega)$, as the inverse of its  largest Lyapunov exponent. This yields
\bea
\label{Lyapunovexponents}
\frac{1}{\xi_{{\rm typ},f}(\omega)}=\lim_{L\rightarrow \infty}\frac{1}{2L}\log\left[\max(\lambda_{1},\lambda_{2})\right],
\eea
where $\lambda_{1},\lambda_{2}$ are the two eigenvalues of ${\cal M}^{T}(\omega){\cal M}(\omega)$.

In the limit of zero energy, $\omega \to 0$, the two blocks of the transfer matrix decouple, and one can immediately read off the typical localization length 
as~\cite{fn2}

\bea
\label{xi_Delta}
\frac{1}{\xi_{{\rm typ},f}(0)}=\left|\overline{\log\left(\frac{\epsilon_{i}}{J}\right)}\right| &\equiv& \left|\log\left(\frac{J_c}{J}\right)\right|\equiv \delta,
\eea
where
\bea
\label{criticalconditionfor1dIsing}
\log(J_c) &\equiv& \overline{\log(\epsilon_{i})}.
\eea

The overbar denotes the disorder average over the random onsite energies 
$\epsilon_i$. $J_{c}$ denotes the critical exchange coupling~\cite{Fisher}, 
at which $\xi_{{\rm typ},f}$ diverges at $\omega=0$, 
and $\delta$ is a dimensionless measure of the distance from criticality.

From Eq.~(\ref{xi_Delta}) one can see that near the critical point, 
the {\em typical} low energy degrees of freedom delocalize as 
$\xi_{{\rm typ},f}(0)\sim \delta^{-\nu}$ with $\nu=1$. 
However, spatially averaged correlation functions are known to decay more slowly,  with a 
faster diverging {\em average} correlation length,~\cite{Fisher} $\xi_{{\rm av},f}(0)\sim \delta^{-2}$. 
This arises because such averages are dominated by rare regions with favorable disorder configuration.

\begin{figure}
\includegraphics[width=3.0in]{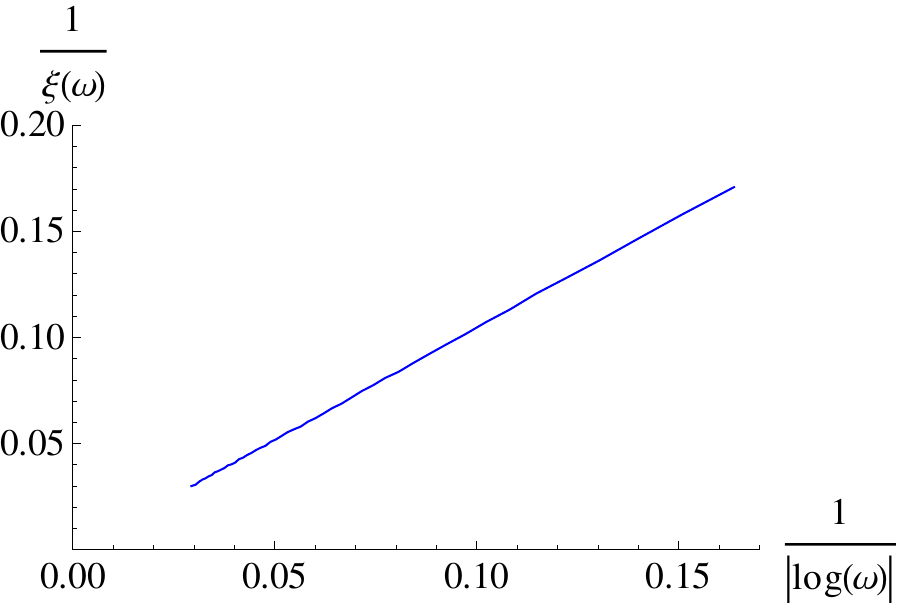}
\vspace{.2cm}
\caption{Lyapunov exponent of the Jordan-Wigner fermions, evaluated numerically from Eq.~(\ref{Lyapunovexponents}), at the critical point of the Ising spin chain, $J=J_{c}$, for box distributed disorder. The localization length exhibits activated scaling $\xi^{-1}(\omega)\propto 1/|\log(\omega)|$.}
\label{f:localizationlengthcritical}
\end{figure}

Note that the typical localization length defined in (\ref{Lyapunovexponents}) is a smooth function of energy $\omega$.
We have studied the energy dependence using the transfer-matrix 
(\ref{transfermatrix}), evaluating $1/\xi_{{\rm typ},f}(\omega)$ numerically 
for a box-distributed disorder $\epsilon_j \in\left[0,1\right]$. 
At the critical point, we find a logarithmically diverging localization length, 
$\xi_{{\rm typ},f}(\omega)\sim |\log(\omega)|$, cf. Fig.~\ref{f:localizationlengthcritical}, 
which is consistent with the activated scaling predicted by the strong randomness renormalization group.~\cite{Fisher}

\begin{figure}
\includegraphics[width=3.0in]{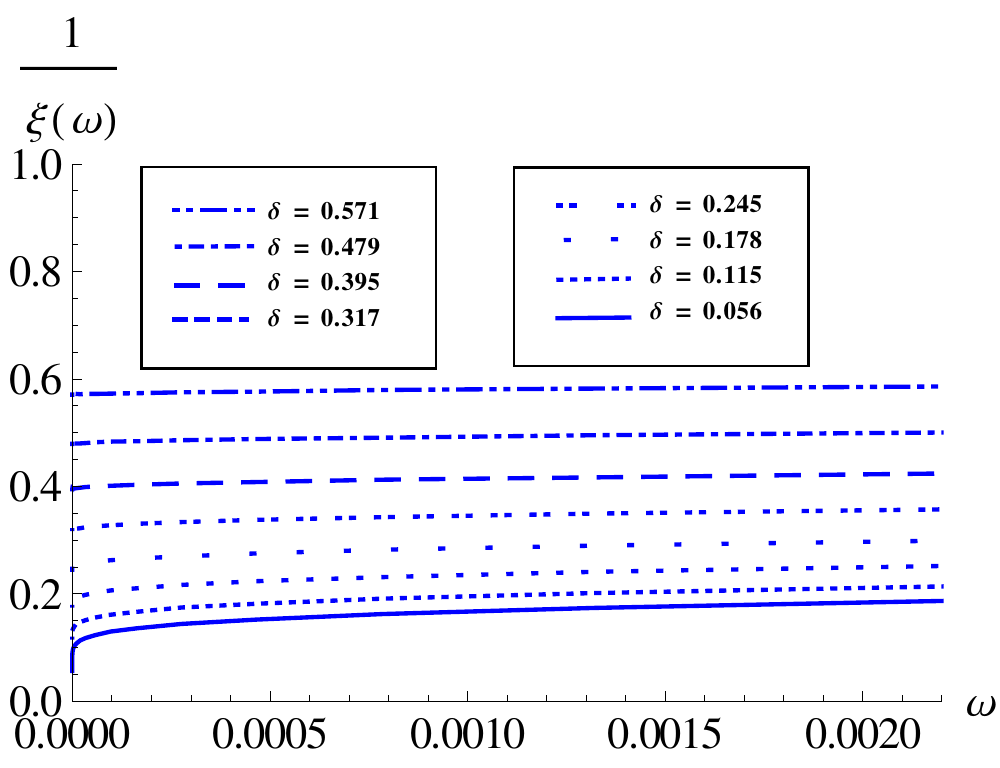}
\vspace{.2cm}
\caption{The numerically evaluated Lyapunov exponent of Jordan-Wigner fermions of the Ising spin chain as 
a function of energy. The disorder is strong, and has a box-distribution in the interval $[-W,W]$ with $W=1$. 
Data is shown for the paramagnetic regime ($J<J_{c}$) off criticality. 
$\delta$ measures the distance from the critical point. At small $\omega$, the localization length $\xi_{{\rm typ},f}$ always {\em decreases} with increasing $\omega$, the leading finite frequency correction being a power law.}
\label{f:localizationlengthoffcritical}
\end{figure}

Away from criticality, the localization length is finite at $\omega=0$, but it behaves non-analytically at small $\omega$
\bea
\label{power1d}
\xi^{-1}_{{\rm typ},f}(\omega)-\xi^{-1}_{{\rm typ},f}(0)\sim \omega^{\alpha},
\eea
 with $\alpha>0$, cf.~Fig.~\ref{f:localizationlengthoffcritical}. We will discuss the origin of this power law and the exponent $\alpha$ in Sec.~\ref{ss:nonanalyticity} below and compare it with exact results obtained in a continuum model.

In the model with box-distributed disorder, at any distance from criticality, we always found the localization length to decrease with increasing energy, all the way  up to the band edge. In other words, the localization length is always an absolute maximum at $\omega=0$.  

\subsection{Continuum limit}
If the disorder is weak, or close to the critical point,
the low energy physics can be captured by coarse-graining the lattice model and 
taking the continuum limit of $\widetilde{{\cal H}}$. 
After rewriting the matrix ${\cal D}$ from (\ref{D}) as
\bea
{\cal D}_{jk}&=&-J(\delta_{j,k-1}-\delta_{jk})+(\epsilon_{j}-J)\delta_{jk},
\eea
the continuum limit of $\widetilde{{\cal H}}$ can be taken as 
\bea
{\cal D}_c=J\left[-\frac{d}{dx}+\phi(x) \right],
\eea
where the lattice spacing was set to unity.
The random potential is given by
\bea
\frac{\epsilon_{j}-J}{J}\to \phi(x),
\eea
where the continuous variable $x$ corresponds to the (coarse-grained) position $j$.

For the case where $\phi(x)$ is a Gaussian white noise potential of unit variance, 
the problem was solved exactly using supersymmetric quantum mechanics.~\cite{Eggarter,Bouchaud,Comtet} 
The continuum version of Eq.~(\ref{Schrodingerequation}) is equivalent 
to the  Schr\"odinger equation for the component $\psi^2$ with the supersymmetric Hamiltonian
\bea
\label{continuummodel}
H_{c}\psi^2&\equiv & {\cal D}_c^\dagger{\cal D}_c \psi^2\\
&=& \left[-\frac{d^{2}}{dx^{2}}+\phi^{2}(x)+\phi '(x)\right]\psi^2 = \omega^2 \psi^2,\nn
\eea
whose spectrum is positive by construction (here and in the remainder of this section we set $J=1$). In Ref.~\onlinecite{Bouchaud} the Lyapunov exponent (inverse localization length) of the eigenfunctions of the continuum Hamiltonian (\ref{continuummodel}) was obtained in closed form as
\bea
\label{exactresult}
\xi_{{\rm typ},f}^{-1}(\omega)=-\omega \frac{dM_{\mu}(\omega)/d\omega}{M_{\mu}(\omega)},
\eea
where $M_{\mu}=\sqrt{J^{2}_{\mu}+N^{2}_{\mu}}$, $J_{\mu}$ and $N_{\mu}$ being Bessel functions of order $\mu$ of the first and second kind, respectively. The index $\mu$ of the Bessel functions is given by the expectation value of the random potential,
\bea
\mu = \overline{\phi(x)}.
\eea

In the vicinity of criticality $\mu$ approaches the value of $\delta$ defined in Eq.~(\ref{xi_Delta}), i.e., $\mu/\delta\to 1$ as $\mu, \delta \to 0$.

Evaluating the expression (\ref{exactresult}) at the critical point, 
$\mu\to 0$, one obtains a logarithmic (activated) scaling at small $\omega$,
\bea
\label{logarithm}
\xi^{-1}_{{\rm typ},f}(\omega)=  \frac{1}{\left|\log(\omega)\right|} +O(1/\left|\log(\omega)\right|^2),   \quad(\mu=0).
\eea

This matches with what we found numerically for the discrete, strong disorder case in the preceding subsection, cf.~Fig.\ref{f:localizationlengthcritical}. 

Close to the critical point, the formula (\ref{exactresult}) predicts 
\bea
\label{xi0}
&&\xi^{-1}_{{\rm typ},f}(0) = \mu,
\eea
and the non-analytic behavior
\bea
\label{power2}
&& \xi^{-1}_{{\rm typ},f}(\omega)-\xi^{-1}_{{\rm typ},f}(0)= A_\mu \omega^{2\mu}[1+o(\omega)],\\
\label{Amu}
&&A_\mu= 2\mu \frac{2^{-2\mu}  }{\Gamma(\mu+1)^2} \frac{\pi \mu}{\tan(\pi \mu)}\,.\nn
\eea
This, too, matches qualitatively the power law behavior we found numerically for 
strongly disordered Ising chains. 

It is interesting to note that in the limits $\mu \rightarrow 0$ and $\omega \rightarrow 0$, but keeping the product $\mu|\log(\omega)|$ constant, the formula~(\ref{exactresult}) can be cast in the following scaling form
\bea
\label{scalingxi}
\xi^{-1}_{{\rm typ},f}(\omega) \to \mu \coth(\mu |\log\omega|)
\eea
which interpolates between the limits of Eqs.~(\ref{logarithm}) and (\ref{power2}). Indeed the latter tends to
\bea
\xi^{-1}_{{\rm typ},f}(\omega) = \mu [1+2 \exp(-2\mu|\log\omega|)+o(\omega)]
\eea
for $\mu|\log(\omega)|\gg 1$.

Let us now proceed to analyze the exponent of the leading power law in $\omega$, and its physical origin in greater generality. 

\subsection{Non-analyticity of $\xi(\omega)$ from rare events}
\label{ss:nonanalyticity}
The origin of the power laws Eqs.~(\ref{power1d},\ref{power2}) can be understood from an analysis of the transfer matrix. This will also elucidate the kind of rare events which lead to the non-analytic low frequency behavior of $\xi(\omega)$.

Let us consider the paramagnetic phase where  $\overline{\log \epsilon_{i}/J}=\delta>0$. 
From Eq.~(\ref{transfermatrix}), it is clear that at $\omega=0$ the Lyapunov exponent, i.e., the growth rate of the norm of the matrix, 
 is given by the product of the elements $T_{22}$. 
For $0<\omega\ll J$, the norm of the transfer matrix is still mostly dominated 
by a product of factors $T_{22}$, but occasionally rare stretches along the chain may occur in 
which locally $\overline{\log \epsilon_{i}/J}<0$. 
If such a rare fluctuation is strong enough it can compensate for the small 
factor $\sim \omega^2$ associated with switching from the $2-$channel to 
the $1-$channel, and thus increases the Lyapunov exponent beyond its $\omega=0$ value.
The increase of $\xi^{-1}$ is proportional to the spatial density of such rare fluctuations. Below we analyze this effect quantitatively.

Consider the product of factors on a stretch of length $\ell$,
\bea
\label{Xl}
X_\ell = \prod_{j=i}^{i+\ell-1} \frac{|\epsilon_j|}{J}.
\eea
For independent, box-distributed $\epsilon_j$ (with $W\equiv 1$) the probability density of the logarithm of $X_\ell$ is easily obtained by a Laplace transform as 
\bea
\label{exactpl}
p_\ell(\log X_\ell=a<a_M)\,da =  \frac{(a_{M}-a)^{\ell-1}}{(\ell-1)!}e^{a-a_{M}}da.\,
\eea
The support extends up to 
\bea
a_M = \ell \log(1/J) = \ell (1+\delta),
\eea
where $\delta$ 
was defined in Eq.~(\ref{xi_Delta}).
For large $\ell$ the exact expression (\ref{exactpl}) can be approximated as 
\bea
\label{approx}
p_\ell(\log X_\ell&=&a)\approx \frac{1}{\sqrt{2\pi \ell}}\times \\
&&\exp\left[-\ell\left(\frac{a_{M}-a}{\ell}-1-\log \frac{a_{M}-a}{\ell}\right)\right].\nn
\eea

For general disorder distributions, one can show that for large $\ell$, up to pre-exponential factors, $p_\ell$ is given by a large deviation expression~\cite{largedeviation}
\bea
p_\ell(\log X_\ell = a) \simeq \exp[-\ell I(a/\ell)].
\eea
Here $I(s)$ is the Legendre transform of the function
\bea
\label{Legendre}
\lambda(k) \equiv \log \left[\overline{ \left(\frac{\epsilon_i}{J}\right)^k }\right], 
\eea
i.e.,
\bea
I(s) = [ks-\lambda(k)]_{k\, :\, \lambda'(k)=s}\,.
\eea

The change in the Lyapunov exponent is given by the sum over contributions from stretches of all lengths, as 
\bea
\label{result}
\Delta \xi^{-1}(\omega) &\propto& 
\int d\ell \,p_{\ell}(\log X_\ell = \log \omega) \\
&\simeq&  \exp[-\ell^* I(\log \omega /\ell^*)] =  \exp[\alpha \log \omega]= \omega^\alpha.\nn
\eea
In the second line the smallness of $\omega$ justifies a saddle point approximation, with the maximal integrand at $\ell=\ell^*\sim |\log \omega|$.~\cite{fnxx} A little algebra shows that the exponent $\alpha$ is given by the strictly positive solution of
\bea
\label{expo}
\lambda(-\alpha)
 =0.
\eea 
Pre-exponential factors of $\log (\omega)$ in Eq.~(\ref{result}) 
can be checked to cancel. Indeed, the normalization factor, 
shown explicitly in the exact expression (\ref{approx}), 
cancels with the contribution from the Gaussian fluctuations around the saddle point taken in Eq.~(\ref{result}). This cancellation of logarithms is also confirmed by the exact result Eq.~(\ref{power2}) for the continuous case.

For a box distributed disorder one finds
\bea
\label{lambdaboxdist}
\lambda(k)&=&-k\log J-\log(1+k)\nn\\
&=& k(1+\delta)-\log(1+k).
\eea
Close to the critical point ($v\ll 1$), the solution of Eq.~(\ref{expo}) yields 
\bea
\label{alpharesult}
\alpha= 2\delta-8 \delta^2/3+ O(\delta^3).
\eea

\begin{figure}
\includegraphics[width=3.0in]{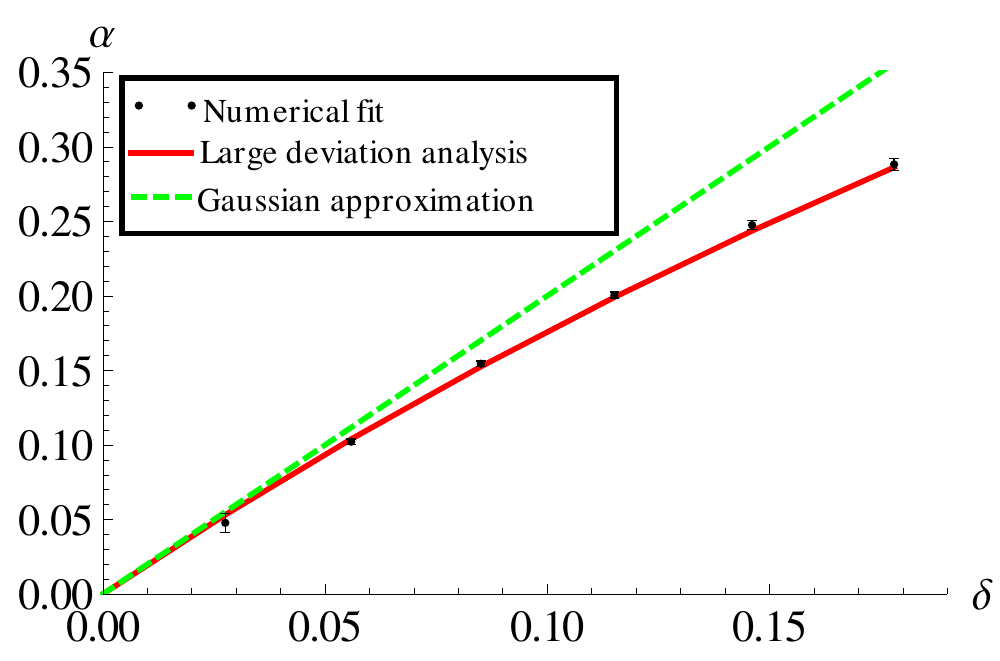}
\vspace{.2cm}
\caption{(Color online) Comparison between the theoretical prediction of Eqs.~(\ref{expo},\ref{lambdaboxdist}) [red solid line] and the numerically obtained exponents $\alpha$ [black dots], extracted from fits of $\xi^{-1}_{\rm typ}(\omega)$ to a power law. 
The abscissa $\delta$ defined in Eq.~(\ref{xi_Delta}), measures the distance to criticality. 
We also show the approximation $\alpha=2\mu/\sigma^2$, obtained by considering $\log(\epsilon_i/J)$ as Gaussian variables, characterized by their mean  $\mu=\overline{\log(\epsilon_i/J)}$ and variance $\sigma^2 = \overline{[\log(\epsilon_i)]^2}-\overline{\log(\epsilon_i)}^2$ (green dashed line).
Close to criticality, this approximation is accurate, while further off criticality large non-Gaussian deviations dominate the behavior of $\alpha$.
}
\label{f:exponents}
\end{figure}

In Fig.~\ref{f:exponents} we show the good agreement between the prediction 
of Eqs.~(\ref{expo},\ref{lambdaboxdist}) with the exponents extracted 
from fits to numerical data for $\xi(\omega)$. 
Note that the result Eq.~(\ref{power2}) of the continuum limit applies close to the critical point, where coarse-graining on large scales is possible. It is recovered in general if $\log(\epsilon_j/J)$ is assumed to be a Gaussian variable with mean $\mu$ and variance $\sigma^2$, in which case $\lambda(k)=\mu k +\sigma^2 k^2/2$ and thus~\cite{fn3} $\alpha =2\mu/\sigma^2$.

The above shows that the leading non-analytic frequency correction to the localization length is due to rare regions which locally favor the minority phase. The results show that low energy excitations are less backscattered from such regions than excitations at higher energy. Furthermore, we see that the regions which dominate the excess backscattering at a given low frequency $\omega$ have a length $\sim|\log \omega|$.

While the above calculation makes a reliable prediction of the exponent in Eqs.~(\ref{power1d},\ref{power2}), the prefactor $A_\mu$ in Eq.~(\ref{power2}) is more sensitive to details of the disorder distribution. As we mentioned before, from the numerical evaluation of the Lyapunov exponents in the strongly disordered spin chains we found $A_\mu$ to be always positive, independently of the distance to criticality. In other words, we find the localization length at $\omega=0$ to be always greater than $\xi_{{\rm typ},f}(\omega>0)$ at higher energies.
However, in the exactly solvable continuum model such a behavior is found only for $\mu<1/2$, cf. Eq.~(\ref{power2}), while further away from criticality the prefactor changes sign. Note that this is not in contradiction with the numerical results for a box distribution, since in the corresponding regime where large deviations become generally strongly non-Gaussian, the two disorder models have no reason to yield similar results.
In contrast, the behavior close to criticality is expected to be well captured by the continuum model with Gaussian disorder, which we have shown to predict the correct exponents $\alpha$ close to criticality. The exactly solvable continuum result then suggests that
close enough to criticality, $\xi_{{\rm typ},f}(\omega)$ assumes a local maximum at $\omega=0$, whatever the specific disorder. Moreover, one may conjecture that the activated scaling form (\ref{scalingxi}) is actually universal close to criticality

\subsection{Implications for spin-spin correlations}
The above results describe the Lyapunov exponents of the free fermions which arise in the Jordan-Wigner decomposition of the 1d Ising chain. 
Their relevance for spin-spin correlation functions is not completely obvious, however. Indeed spin correlation functions, which inform about the localization properties of spin excitations, are difficult to extract in general from the fermion representation, because of the nonlocal relation between spin and fermion operators.
Nevertheless, we expect that the localization length of fermion Green's functions also controls the spatial decay of spin correlations at low enough energies.

It is interesting, in particular, to compare the exact results for the free fermions to the results we obtained to leading order for the spin problems. Applying the locator expansion to the spin chain, and defining the localization length via the spin correlation function as ~\cite{positivemagnetoresistance}
\bea
\label{exactLyapunovexponent}
\xi^{-1}_{s}(\omega)=-\lim_{l \rightarrow \infty}\frac{\overline{\log(|G_{l,0}(\omega)/G_{0,0}(\omega)|)}}{| l |},
\eea
we find from Eq.~(\ref{Isingmatrixelement}) the expression
\bea
\label{Lyapunovexponentslocaterexpansion}
\xi^{-1}_{s}(\omega)=-\lim_{l\rightarrow \infty}\frac{1}{l} \sum^{l}_{j=1}\overline{ \log \left[
\frac{4J |\epsilon_{j}|}{(2\epsilon_{j})^{2}-\omega^{2}}\right]},
\eea
to leading order in $J$. Remarkably, at zero frequency $\omega=0$, the exact result for JW fermions, Eq.~(\ref{xi_Delta}) and the locator expansion (\ref{Lyapunovexponentslocaterexpansion}) yield the same result. This suggests that at least at low energies fermion localization lengths are indeed good indicators for spin-spin correlation lengths.

\subsection{Protected resonances at $\omega=0$ in general Ising models}
\label{ss:resonances}
The above implies that the leading order locator expansion is actually {\em exact} at $\omega=0$.
The reason for this phenomenon is that in Eq.~(\ref{Schrodingerequation}), when $\omega=0$, 
the fermion modes $\psi^1$ and $\psi^2$ decouple and satisfy independent equations, 
which are easily solved by forward integration. Thus, no higher order corrections from loops arise.~\cite{fn4}

The fact that, at $\omega=0$, the denominators $1/\epsilon_i$ are not renormalized by higher order corrections (such as self-energies, as in the standard Anderson model) is deeply rooted in the Ising symmetry.
To see this, let us suppose that one of the transverse fields vanishes, $\epsilon_i=0$.
The Ising Hamiltonian always enjoys the discrete {\em global} Ising (parity) symmetry, 
\bea
U_{\rm Ising}=\prod_j \sigma_j^z,
\eea
satisfying $U_{\rm Ising} H U_{\rm Ising}^{-1}=H$. However, if $\epsilon_i=0$ it possesses the  {\em additional local} symmetry,
\bea
U_{\rm loc} = \sigma_i^x,
\eea
with $U_{\rm loc} H U_{\rm loc}^{-1}=H$. Since $U_{\rm Ising}$ and $U_{\rm loc}$ do not commute, it immediately follows that 
every eigenenergy of the full Hamiltonian is twofold degenerate. In particular the ground state is doubly degenerate.
 Note that this conclusion is valid {\em independently} of the dimension.

It is interesting to note that in the paramagnetic phase the two degenerate ground states differ only in observables depending 
on degrees of freedom close to the site $i$, as one can explicitly show in the case of a 1d chain.
As a consequence of this local degeneracy, correlation functions evaluated in the ground state  will be singular at $\omega=0$. 
This is only ensured if the pole of the bare denominator $1/\epsilon_i$ never shifts away from $\epsilon_i=0$ 
by any higher order exchange corrections, when evaluated at $\omega=0$. 
In other words there cannot be any finite self-energy-like corrections at $\omega=0$ in the ground state of the Ising model, 
as an onsite energy approaches $\epsilon_i\to 0$. This contrasts with the XY model,
 where there is no symmetry reason that suppresses exchange corrections to resonant small denominators.
 Those self-energy-like corrections actually play a crucial role in regularizing correlation functions and localization properties. 
We will discuss the consequences of protected resonances at $\omega=0$ on the nature of the ordering transition further below in Sec.~\ref{s:delocalization}.

\subsection{Localization length at finite $\omega$: $d=1$}
At finite $\omega$, the leading order locator expansion in $d=1$ captures correctly the qualitative feature that $\xi^{-1}_{s}(\omega)$ decreases with increasing energy. However, it misses the non-analytic corrections  due to rare regions whose disorder strength is typical for the opposite phase than realized in the bulk of the sample. Those regions lead to backscattering, which tends to enhance the localization of higher energy excitations. 
In the next section we will discuss the effect of such rare regions on the Cayley tree, and argue that, similarly as in 1d, there are non-analyticities in $\xi(\omega)$ at $\omega=0$ off criticality, and activated scaling at criticality. 

We mention in passing that the non-analyticity of the localization length in 1d, 
as well as its divergence at the critical point in 1d, 
come along with an accompanying Dyson singularity in the density 
of fermionic states,~\cite{Bouchaud} very similarly as in tight-binding chains 
with off-diagonal disorder.~\cite{Dyson,Eggarter,Balents} 
Both occur naturally due to the BDI symmetry of the fermionic problem.

\subsection{Localization length at finite $\omega$: $d>1$}
In higher dimensions, $d>1$, there is a further reason for the localization length to decrease with increasing frequency, as discussed in Ref.~\onlinecite{Markus}: The interference between alternative forward directed paths in Eqs.~(\ref{Isingmatrixelement},\ref{XYmatrixelement}) is maximally constructive at vanishing excitation energy $\omega=\epsilon_0=0$, while at finite $\omega$ negative scattering amplitudes arise, which spoil the perfect interference and decrease the propagation amplitude at large distances.  

Qualitatively similar effects are achieved by a magnetic field acting on charged bosons, which endows the various paths with different Aharonov-Bohm phases, which also degrade the perfect interference. The resulting magnetoresistance was discussed in detail in Ref.~\onlinecite{Gangopadhyay2012}.

The arguments for both these effects rely a priori on the lowest order expansion in the exchange $J$. However, we expect those to capture the essential features of localization in $d>1$ well within the strongly localized regime. Subleading effects due to loop corrections will be discussed in more details elsewhere.~\cite{BapstMueller} Closer to criticality and at sufficiently high energies, rather high orders of the exchange expansion may become relevant. The above conclusions should thus not be applied without caution to that regime. 

\section{Approaching delocalization: Boson and spin models on highly connected Cayley tree}
\label{s:delocalization}
In an attempt to approach the delocalization transition, 
we now apply our formalism to a situation where the expansion in exchange 
is expected to remain applicable even close to the phase transition. 
A priori one expects this to be the case in high dimensions, 
where subleading loop corrections can be expected to be relatively unimportant. 
An extreme case where loops are absent altogether is the Cayley tree. 
Motivated by closely related studies~\cite{FeigelmanIoffeMezard,IoffeMezard} 
we consider Cayley trees of large branching number $K$ which locally resemble cubic lattices 
in $d=(K+1)/2$ dimensions. The related Anderson model of non-interacting fermions on such trees 
can be solved exactly due to the absence of loops.~\cite{Abou-Chacra, Aizenman}
Since it is known that in this case the delocalization transition happens 
when the hopping is still parametrically small, 
$t_c\sim 1/K\log(K)$ as $K\to\infty$, one may hope that a leading order 
expansion in the exchange/boson hopping can capture the vicinity of the transition.
Such an approach was proposed in Refs.~\onlinecite{FeigelmanIoffeMezard,IoffeMezard} 
where the localization properties of intensive low energy excitations 
in the disordered phase were studied. 
The authors claimed that in the disordered regime, 
close enough to criticality an intensive mobility edge $\omega_c$ exists, that separates delocalized high energy excitations from localized low energy excitations.
Furthermore, it was stated that upon approaching criticality $\omega_c$ decreases to zero and 
vanishes simultaneously with the onset of long range order. 
Similar scenarios had been proposed by other authors as well.~\cite{HertzAnderson,Markus} 
Here we revisit this question, using the leading order locator expansion formulas (\ref{Isingmatrixelement},\ref{XYmatrixelement}). However, they differ from the expressions postulated in Refs.~\onlinecite{FeigelmanIoffeMezard,IoffeMezard} in crucial details, which lead us to qualitatively different conclusions.

\begin{figure}
\includegraphics[width=3.0in]{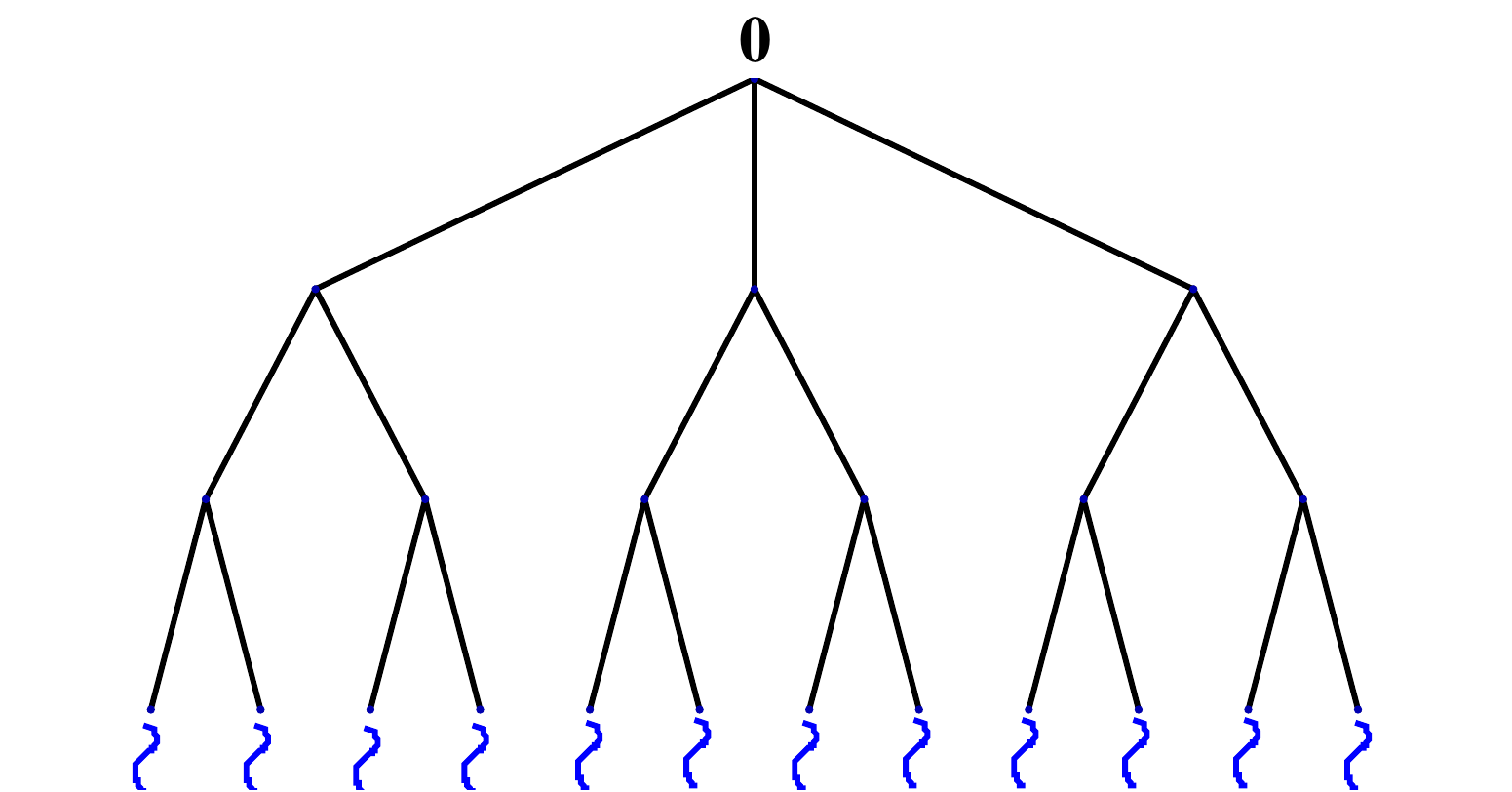}
\vspace{.2cm}
\caption{Cayley tree with branching number $K=2$. The depth of the tree, i.e., the distance from the root to a boundary site is $L$ (in this figure $L=3$).
The wavy lines represent the coupling of the boundary spins to independent baths. }
\label{f:Cayleytree}
\end{figure}

Let us now analyze the spin models Eqs.~(\ref{disorderedXY}, \ref{disorderedIsing}) on a Cayley tree with a root site $0$, branching number $K$ and depth $L$, cf. Fig.~\ref{f:Cayleytree}.
We will be interested in two distinct aspects: (i) the propagation of long range order (emergence of \textquotedblleft surface magnetization\textquotedblright), (ii) the localization properties of local excitations at the root, as a function of the associated energy $\approx 2\epsilon_0$.

Anticipating that the delocalization transition appears when  the exchange is of order $J\sim 1/K\log(K)$, as for single particle hopping, we introduce the notation 
\bea
J \equiv \frac{g}{K}.
\eea
We then rely on the smallness of the parameter $J$ close to criticality to restrict ourselves to leading order perturbation theory in $J$. This will give indeed a reliable estimation of localization properties up to a narrow critical window close to the phase transition, where subleading terms should be included, as we will discuss in detail below.

\subsection {Ordering transition out of the paramagnetic phase}

The disorder-induced quantum phase transition in the models~(\ref{disorderedXY}, \ref{disorderedIsing}) 
can be approached from the ordered or the disordered side.  
The route from the symmetry broken side was pioneered by Ioffe, M\'ezard and Feigelman,~\cite{FeigelmanIoffeMezard,IoffeMezard} 
where (approximate) self-consistent equations for cavity mean fields (local order parameters) were analyzed and solved. 
This inhomogeneous mean field approach was further exploited by Monthus and Garel,~\cite{MonthusGarel} both in finite dimensions 
and on the Cayley trees, finding that strong randomness physics governs the excitations in these disordered systems.

These works pointed out the close relationship between magnetic correlation functions at large distances and the physics 
of directed polymers in random media, which may indeed play a role in the distribution of local order parameters, as recent experiments  in strongly disordered 2d superconductors suggest.~\cite{Castellani}
The mapping between bosonic correlation functions and directed polymers is 
rendered exact on the insulating side $J\ll1$ where the correlation functions can be argued to be well 
represented by the lowest order expansion in the exchange.~\cite{positivemagnetoresistance,Gangopadhyay2012}  
By restricting to the leading order the expansion of order parameter correlations in the ordered phase, or of 
two point functions like $G_{l0}$ in the localized phase, one obtains the same estimate for the critical 
point $J_c=g_c/K$ where long range correlations set in. However, it is hard to assess the quality of the mean field approximation,
 and to improve systematically beyond it, which would be desirable especially in low dimensions. 
In contrast,  higher order corrections in the localized phase are amenable to a systematic expansion in powers of $J$, 
and thus seem a simpler route towards describing criticality.

In this section we approach the ordering transition from the insulating phase. We define the surface susceptibility, i.e., the susceptibility to a homogeneous field $h_x$ applied to the boundary sites, as
\bea
\chi_{s}\equiv \sum_{l\in \partial \Lambda} \frac{G_{l,0}(0)}{G_{0,0}(0)}.
\eea
The sum is over all boundary spins $l$, 
and we included a normalization factor $G_{0,0}(0)$ for convenience. 
In the insulating phase, all $G_{l,0}$ decay rapidly, 
such that the large number of boundary sites cannot offset the smallness of this susceptibility.
Upon increasing the exchange coupling, $g$, the ordering transition
(in typical realizations of disorder) occurs when the {\em typical} value of  
$\chi_{s}$ is of order $O(1)$. 
In finite dimensions this criterion is equivalent to asking 
that $-\log G_{l,0}$ does not grow linearly with the distance 
from the bulk site $0$ to a boundary site $l$. 
However, on the Cayley tree, where there are exponentially many ($K^L$) boundary sites, 
the criterion of non-exponential decay with $L$ must be applied to the whole sum, and cannot so easily be reduced to a criterion on typical or dominant paths on the tree.

The above criterion is valid on general lattices. However, in finite dimensions, 
the Green's functions $G_{l,0}$ are hard to study analytically, especially close to the phase transition. 
In contrast, on a Cayley tree with large branching number $K$ a simplification occurs. 
First of all, there is only one shortest path connecting any two points, 
and thus there is only a single term contributing to the leading order locator expansion of $G_{l,0}$.
Subleading terms are not very important when $K$ is large, since paths with extra excursions on side paths 
are formally penalized by an extra factor of $\lesssim g_c^2/K$ (which is dominated by exchange processes 
with the most favorable neighboring sites). This argument is however known to be a bit too naive. 
Indeed, from the analogous single particle problem,~\cite{Anderson,Abou-Chacra} 
it is known that these self-energy corrections regularize resonances from very small denominators and modify
the numerical prefactor $A$ in the large $K$ scaling for the critical hopping, $g_c= A/(K\log K)$, as compared to the so-called
\textquotedblleft Anderson upper limit \textquotedblright estimate, 
in which self-energy contributions are neglected. Similar effects are {\em a priori} to be expected in the many body case as well.

Keeping this caveat in mind, we nevertheless start by restricting ourselves to the leading order perturbation theory. We use the results of Eqs.~(\ref{Isingmatrixelement}) or (\ref{XYmatrixelement}) to evaluate $G_{l,0}$ as
\bea
\label{chis}
\chi_{s}=\sum_{l\in \partial \Lambda} \prod_{i\in {\cal P}_l}\frac{g/K}{|\epsilon_{i}|},
\eea
where ${\cal P}_l$ is the unique path from the root to the boundary site $l$.
This certainly captures well the behavior deep in the insulator, 
but as argued above, also rather close to the ordering transition if 
the limit of large $K\gg 1$ is taken. 
We recall that to this leading order the two point functions at $\omega=0$ are 
the same in the XY and the Ising model, which thus leads to the same estimate of the critical coupling $g_c$~\cite{IoffeMezard, FeigelmanIoffeMezard}

A similar conclusion was reached based on cavity mean field equations applied to the ordered side, linearizing them in the exchange coupling and the local order parameter.
As the local order parameter susceptibility is $1/\epsilon_i$ in both models, the mean field approximation estimates the order parameter susceptibility at large distances to be a product of local susceptibilities, resulting in the same expression as Eq.~(\ref{chis}) derived for the insulating side - independently of the symmetry of the order parameter.

The (near) coincidence of the critical values $g_c$ in the two models appears less obvious when reasoning from the disordered side. Naively, one might think that the additional exchange term $J \sigma_i^y \sigma_j^y$ in the XY model leads to enhanced fluctuations as compared to the Ising model. However, this effect is almost exactly compensated by the fact that the XY symmetry (conservation of hard core bosons) restricts the quantum fluctuations more strongly than the Ising symmetry.
   
However, we will argue below that the coincidence of the two values of $g_c$ is an artifact of the various approximations. In fact, the inclusion of subleading terms in the locator expansion will be shown to split the degeneracy of the two critical values, see Eq.~(\ref{gcs}). Since these corrections regularize resonances (small denominators) in the XY model, they have a non-perturbative effect that modifies $g_c$ at the {\em leading order} in the large $K$ asymptotics. For the time being we nevertheless carry on with the analysis to the leading order in exchange, which will set the stage for later refinements. 

\subsection{Mapping to directed polymers}
\label{ss:dirpol}
The evaluation of the typical susceptibility can be accomplished via the exact mapping of the leading order expression for the surface susceptibility to the problem of a directed polymer on the Cayley tree. The latter was solved exactly by Derrida and Spohn,~\cite{directedpolymer} and their result was applied to the present context in Refs.~\onlinecite{IoffeMezard, FeigelmanIoffeMezard}.
The susceptibility itself is a strongly fluctuating random variable, which depends on the disorder realization. However, its logarithm is a self-averaging quantity. Upon re-exponentiation one obtains the {\em typical} value, which is characterized by the logarithmic disorder average~\cite{directedpolymer,IoffeMezard, FeigelmanIoffeMezard}
\bea
\label{chityp}
\lim_{L\rightarrow \infty} \frac{1}{L}\overline{\log\chi_{s}}&\equiv& \log\left(\frac{g}{K}\right)+\min_{x\in[0,1]}f(x),
\eea
where the function $f(x)$ is defined by
\bea
\label{targetfunction}
f(x)=\frac{1}{x}\log\left[K \int^{1}_{-1} \frac{1}{\mid \epsilon \mid^{x}} \frac{d\epsilon}{2} \right].
\eea
Let us denote by $x_{c}$ the argument at which $f(x)$ takes its minimum on the interval $x\in [0,1]$. If $x_c<1$, the associated directed polymer problem is in its low temperature frozen phase whose thermodynamics is essentially dominated by a single path on the tree. More precisely, the partition sum over paths that go to the boundary is dominated by a set of configurations, which all stay together and split only at a short distance before reaching the boundary. Thereby, that last distance does not scale with $L$ in the limit $L\to \infty$. In spin glass terminology, the dominating paths have \textquotedblleft mutual overlap\textquotedblright tending to $1$ in the thermodynamic limit. This situation corresponds to a phase of broken replica symmetry (RSB) for the directed polymer. In the language of onsetting long range order in the spin model this translates into the statement that (in leading order approximation in the exchange) the surface susceptibility is dominated essentially by one or a few single paths. 
If instead one finds $x_{c}=1$  the
dominant contribution to $\chi_{s}$, or to the partition function of the equivalent polymer problem, is due to infinitely many configurations in the thermodynamic limit. We will come back to these issues in Sec.~\ref{s:RSB} where we will discuss the difference between Ising and XY models in this respect.

Within the approximation to leading order in the hopping one finds that the ordering transition occurs when,~\cite{IoffeMezard}
\bea
0=\log\left(\frac{g_c}{K}\right)+f(x_{c}),
\eea
which corresponds to the vanishing of the free energy of the directed polymer per unit length. This has the solution~\cite{IoffeMezard}
\bea
\label{SItransition}
g_{c}\exp\left(\frac{1}{eg_{c}}\right)=K, \quad x_{c}=1-eg_{c},
\eea
or, in the limit of large $K\gg 1$,
\bea
\label{naivegc}
g_{c}\approx \frac{1}{e\log(K)}.
\eea
It is worth noting that the Eqs.~(\ref{chityp},\ref{targetfunction}), 
which determine the critical exchange $g_c$, are identical to those 
obtained by Abou-Chacra et al.~\cite{Abou-Chacra}  for the delocalization of non-interacting 
particles, within the so-called \textquotedblleft Anderson upper limit\textquotedblright~approximation. The latter consists in dropping self-energy corrections, which is equivalent to the leading order approximation in hopping.~\cite{Anderson}
The coincidence of these results is not very surprising, since the localization properties of fermions and hard core bosons are very similar on the Cayley tree. In fact, to leading order in hopping, one considers only forward scattering processes, and since the Cayley tree does not contain loops, the quantum statistics of the particles is irrelevant at that order. Similarly, to leading order in the hopping, the dependence of localization properties on frequency will not differ between fermions and hard core bosons, as we will see in the following subsection.

\subsection {Spatial decay rate in the paramagnet phase of the XY model}
Let us now turn to the localization properties in the insulating phase ($g<g_{c}$), where the locator expansion is best controlled.
We are interested in particular in determining whether there exists an intensive mobility edge, i.e., an energy of order $O(1)$  which separates localized from delocalized excitations in the many body system, as claimed in Refs.~\onlinecite{IoffeMezard,FeigelmanIoffeMezard}.

In order study the decay process of a local excitation on the root $0$, we couple our system to zero temperature baths via the spins at the boundary  $\partial\Lambda$ of the Cayley tree, cf. Fig.~\ref{f:Cayleytree}. %In terms of the general formalism of Sec.~\ref{s:decay}, we take the Cayley tree as the lattice $\Lambda$, and its leaves as the boundary set. 
On a Cayley tree, there is only one shortest path between the root site $0$ and any boundary site $l$. This simplifies the analysis of decay rates significantly, since to leading order no path interferences need to be taken into account. As discussed above, for $K\gg 1$ the leading  order approximation captures rather well the insulating phase, because it disfavors subleading corrections in $g/K$, which arise from paths with transverse excursions.
Within this approximation we evaluate the decay rate of a local excitation with energy $\omega\simeq 2\epsilon_{0}$ as
\bea
\label{levelwidth}
\Gamma_{0}(\omega)=\sum_{l\in \partial \Lambda} \prod_{i\in {\cal P}_l} \left[\frac{2g/K}{2\epsilon_{i}-\omega}\right]^{2}J_b(\omega).
\eea

As in the calculation of the surface susceptibility, the decay rate $\Gamma_{0}(\omega)$ can be seen as the partition function for a directed polymer on the tree, whereby the squares of the locators $2g/K(2\epsilon_i-\omega)$ take the role of random local Boltzmann weights.

The typical value $\Gamma_{0}$ at fixed frequency $\omega$ is best characterized by its mean spatial rate of decrease,
\bea
\gamma_{\rm XY}(\omega)&\equiv&-\lim_{L\rightarrow \infty} \frac{1}{L}\overline{\log\left[\frac{\Gamma_{0}(\omega)}{J_b(\omega)}\right]}\nn\\
&=& -\left[\log\left(\frac{g}{K}\right)+\min_{x\in[0,1]} f_{\omega}(x)\right],
\eea
where the function $f_{\omega}(x)$ is defined by
\bea
\label{targetfunctionXY}
f_{\omega}(x)=\frac{1}{2x}\log \left(K \int^{1}_{-1} \frac{1}{\mid \epsilon-\omega/2 \mid^{2x}}\frac{d\epsilon}{2} \right).
\eea
Notice that due to the assumed symmetry of the onsite disorder distribution, we have $f_{-\omega}(x)=f_{\omega}(x)$, so it suffices to study $\omega>0$. Suppose again that $f_{\omega}(x)$ takes its minimum on the interval $[0,1]$ at $x=x_{c}$. Due to the small denominators arising from sites with $\epsilon_i\to 0$, Eq.~(\ref{targetfunctionXY}) is well-defined only for $x<1/2$ and thus $x_{c}$ must be less than $1/2$. 
Further below we will discuss however that  this restriction does not apply when the resonant levels are regularized by self-energy corrections.

\subsubsection{Search for a mobility edge}
\begin{figure}
\includegraphics[width=3.0in]{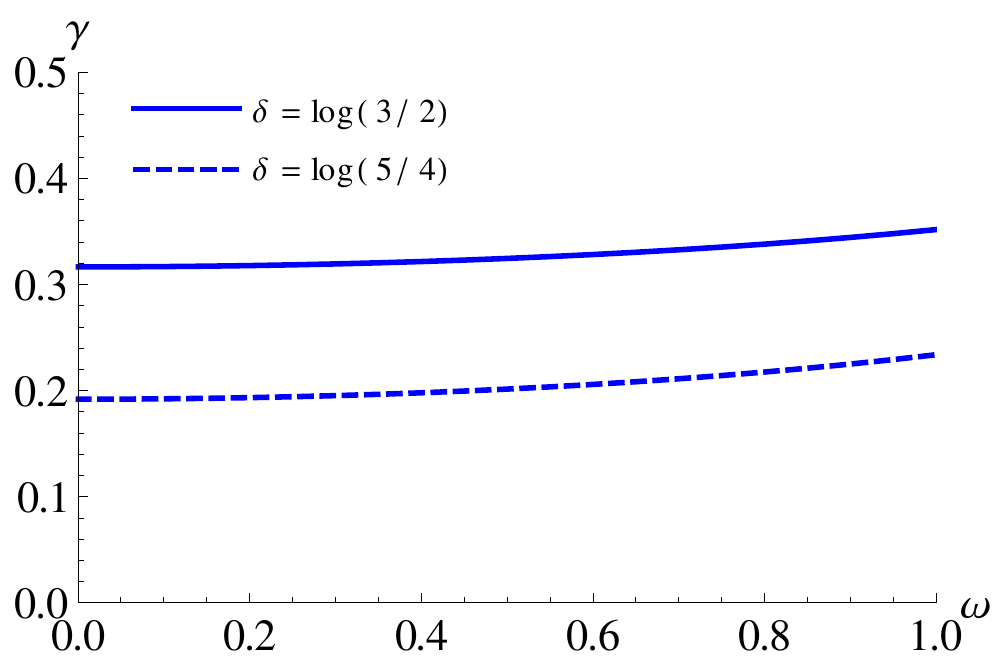}
\vspace{.2cm}
\caption{
Spatial decay rate of excitations in the XY model as a function of frequency
$\omega$. The two curves correspond to exchange constants $g/g_c=2/3$ and $4/5$ ($\delta=\log(3/2)$ and $\log(5/4)$), respectively, evaluated on a Cayley tree of
connectivity $K=2$. The lowest frequency excitations have the slowest decay rate with spatial distance. 
 }
\label{f:DecayratefortheXYmodel}
\end{figure}

 $\gamma_{\rm XY}(\omega)$ controls the decay or growth of $\Gamma_{0}(\omega)$ with $L$. Clearly, as long as $\gamma_{\rm XY}(\omega)>0$, the typical value of $\Gamma_{0}(\omega)$ is exponentially small as $L\rightarrow\infty$,
which implies that an excitation of energy $\omega$ is localized and does not decay into the bath in the thermodynamic limit $L\rightarrow \infty$. Instead, the delocalization of an excitation with energy $\omega$ occurs at the point where
\bea
0=\gamma_{\rm XY}(\omega)=\log\left(\frac{g}{K}\right)+\min_{x\in[0,1/2]}f_{\omega}(x).
\eea
By minimizing the function $f_{\omega}(x)$ with respect to $x$, one obtains the two simultaneous conditions for $\omega$ and $x=x_\omega$, 
\bea
\label{rsxy}
1=\frac{K}{2}\left(\frac{g}{K}\right)^{2x_\omega}\int^{1}_{-1}\frac{1}{|\epsilon-\omega/2|^{2x_\omega}}d\epsilon,
\eea

\bea
\label{minimumxy}
\log\left(\frac{g}{K}\right)=\frac{K}{2}\left(\frac{g}{K}\right)^{2x_\omega}\int^{1}_{-1}\frac{\log(|\epsilon-\omega/2|)}{|\epsilon-\omega/2|^{2 x_\omega}}d\epsilon.
\eea
However, one finds that on the disordered side of the transition, $g<g_{c}$, there is no $\omega>0$ such that Eqs.~(\ref{rsxy}) and (\ref{minimumxy}) are satisfied simultaneously. In other words, at this order of perturbation theory there is no indication of a critical energy (mobility edge) above which excitations
are  delocalized in the insulator. Rather, the excitations with intensive energy are all localized in the quantum paramagnet (Bose insulator).
Moreover, we find that in the whole localized phase, $g<g_{c}$, 
$\gamma_{\rm XY}(\omega) >\gamma_{\rm XY}(0)$ for any $1\geq |\omega|>0$, cf. Fig.~\ref{f:DecayratefortheXYmodel}.  

\subsubsection{Comparison with earlier studies}
The above results contradict those reported in Refs.~\onlinecite{FeigelmanIoffeMezard,IoffeMezard}. On a technical level, the difference between the two results arises because in Refs.~\onlinecite{FeigelmanIoffeMezard,IoffeMezard} the authors restricted the on-site energies $\epsilon_{i}$ to be positive, and postulated a matrix element of the form $\prod_i \left[\frac{2g/K}{2|\epsilon_{i}|-\omega}\right]^{2}$ as in Eq.~(\ref{levelwidth}). However, the former can be imposed only for the Ising model without restricting generality. On the other hand, the Ising model requires the use of a different matrix element than in (\ref{levelwidth}), see Eq.~(\ref{levelwidth2}). Hence, the actual calculation neither applied to the XY nor to the Ising model.   

As the results predicted from that calculation lead to qualitatively different results from ours, it is instructive to analyze in more detail the reasonings and pitfalls which may lead to it. The above form of the matrix element was probably guessed from perturbation theory where spins are flipped progressively along the path - even though this yields a decay rate which is $2^L$ times smaller than Eq.~(\ref{levelwidth}). Such a guess can be motivated by viewing spin flip propagation as a linearly progressing, single-particle like process. However, in reality the process is much more complex, because virtual intermediate states with many spin flips contribute as well. A flavor of this  may be obtained from the non-trivial example treated in the Appendix via standard perturbation theory.
 
A similar thinking underlied the reasoning in Ref.~\onlinecite{Markus} where it was argued that excitations close to the chemical potential should be more localized than at higher intensive energies. The idea was that such excitations behaved as if they were at a band edge of a single particle problem, given that any local excitation costs a positive energy. For an Ising model, this idea is equivalent to restricting virtual states of the perturbation theory to states with only one single spin flip, and thus leads to the incorrect conclusion that the localization length should always increase with increasing energy. A similar reasoning for XY models or hard core bosons would impose an artificial restriction of perturbation theory to intermediate states where only one extra boson is allowed to be placed on a formerly empty site, with no rearrangements of other particles. However, this is is incorrect since it neglects the indistinguishability of particles and the resulting exchange effects. 
This is seen most easily by considering a non-interacting disorder Fermi sea (Anderson insulator), where an analogous restriction of perturbation theory 
would lead to the obviously incorrect conclusion that the excitations are most localized at the Fermi level.

\subsubsection{Decay rate close to criticality}
Let us now study the behavior of $\gamma_{\rm XY}(\omega)$ for small $\omega$ close to the critical point, within the leading order perturbation theory. For $0< \omega \ll 1$, we may expand
\bea
f_{\omega}(x)=\frac{1}{2x}\log \frac{K}{1-2x}-\left(\frac{1}{2}-x \right)\left(\frac{\omega}{2}\right)^{2}+o(\omega^2).\,\,
\eea
Using this close to the critical point, we deduce the frequency dependence of the spatial decay rate as 
\bea
\label{XYdecayrate}
\gamma_{\rm XY}(\omega; g=g_c)&=& -\left[\log\left(\frac{g}{K}\right)+\min_{x\in[0,1/2]}f_{\omega}(x)\right]\nn\\
&\simeq& \delta+\left(\frac{1}{2}-x_c\right)\left(\frac{\omega}{2}\right)^2,\,
\eea
where we recall the definition $\delta = \log\left( J_{c,{\rm XY}}/J\right)=\log\left( g_{c,{\rm XY}}/g\right)$, measuring the distance to criticality.
Here the parameter $x_c$ satisfies $df_{\omega=0}/dx(x=x_c)=0$. At criticality this becomes
\bea
\label{xcXY}
\log \left(\frac{g_c}{K}\right)+ \frac{1}{1-2x_c}=0,
\eea
upon using the condition of criticality, $\log\left(\frac{g_c}{K}\right)+f_{\omega=0}(x_c)=0$. 
However, in the next subsection, Eq.~(\ref{gammaXY}), when taking into account the most important subleading corrections, we will find a coefficient of the term $\sim\omega^2$ which is differs from Eq.~(\ref{XYdecayrate}). 

\subsection{Self-energy corrections in the XY model}
Notice that Eqs.~(\ref{rsxy}) and (\ref{minimumxy}) are identical to the condition for a single particle delocalization transition 
on a Cayley tree, if the real parts of the local self-energies are neglected in that problem.~\cite{Abou-Chacra} 
It is well-known~\cite{Anderson} that the main physical effect of those self-energies is to moderate the influence of strong 
resonances due to sites with $\epsilon_i\approx \omega/2$. 
Those lead to large self-energies $\Sigma(\omega) \approx J^2/(\epsilon_i-\omega/2)$ on the neighboring sites, 
producing a large denominator in the locator expansion, which tends to neutralize the effect of the resonance. 
As we will recall below the inclusion of this effect corrects the location of the transition point by 
a factor $e/2$ in the limit of large $K$, but is of little further consequence for delocalization.~\cite{Anderson, Abou-Chacra, Thoulessreview}
A simple, but quite accurate way to take into account such self-energy effects, which arise in higher order of perturbation theory,~\cite{BapstMueller} 
is sometimes referred to as \textquotedblleft Anderson's best estimate\textquotedblright. It consists in modifying the local density of states by excluding sites with energies closer $\omega/2$ than a distance 
\bea
\Delta\equiv J^2= (g/K)^2, 
\eea
since those tend to self-neutralize. This leads to a modification of the function $f_\omega(x)$ as 
\bea
f^\Delta_{\omega}(x)=\frac{1}{2x}\log\left[K \int \frac{\rho_{\Delta}(\epsilon)}{\mid \epsilon-\omega/2 \mid^{2x}} d\epsilon \right].
\eea
Here $\rho_{\Delta}(\epsilon)\equiv \Theta(|\epsilon-\omega/2|-\Delta)\rho(\epsilon)$ simply excludes paths through sites with strong resonances. 
With this modification one finds that $f^\Delta_{\omega=0}(x)$ is minimized by $x_c=1/2$ and thus 
\bea
&&\min_{x\in[0,1/2]} f^\Delta_{\omega}(x)= \log K\\
&& \quad\quad\quad
 +\log\left[2\log\left(\frac{K}{g}\right) -\frac{1}{2}\left(\frac{\omega}{2}\right)^2+o(\omega^2)\right].\nn
\eea

Using that $df^\Delta_{\omega=0}/dx(x_c=1/2)=0$, this approximation gives the modified condition for delocalization (critical exchange $g_c$) at $\omega=0$ 
\bea
K= g_{c,{\rm XY}}\exp\left( \frac{1}{2g_{c,{\rm XY}}}\right).
\eea
At large $K$, this tends to
\bea
\label{gcXY}
g_{c,{\rm XY}} \approx \frac{1}{2}\frac{1}{\log K},
\eea
which modifies Eqs.~(\ref{SItransition},\ref{naivegc}). For single particles the above results have been rigorously proven to give the correct leading asymptotics at large connectivity $K$.~\cite{Abou-Chacra, Aizenman, Victor2013} Considering that the leading terms and the dominant subleading corrections are the same for hard core bosons, it is very likely that the same leading asymptotics holds rigorously for hard core bosons as well.~\cite{BapstMueller}

Close to the critical point, $g=g_{c,{\rm XY}}$, we extract from the above the frequency-dependent decay rate
\bea
\label{gammaXY}
\gamma_{\rm XY}(\omega) 
\simeq \delta +\frac{1}{4\log(K/g_{c,{\rm XY}})}\left(\frac{\omega}{2}\right)^2.
\eea 
Note that at large $K$ the prefactor scales as $1/\log K$, as also predicted by the calculation that neglected self-energy effects, cf. Eqs.~(\ref{XYdecayrate}, \ref{xcXY}). However, the numerical prefactor is different.
We will analyze the the range of validity of Eq.~(\ref{gammaXY}) further below.

\subsection {Spatial decay rate in the Ising paramagnet to leading order in $J$}

Let us now analyze the Ising model in turn and contrast it with the XY model.
The relevant matrix element $\langle\textrm{GS}|\sigma^{x}_{l}|E_{0}\rangle$ for the Ising model is given by Eq.~(\ref{matrixelement}). Hence, the decay rate of a local excitation with energy $\omega$ at site $0$ can be written as
\bea
\label{levelwidth2}
\Gamma_{0}(\omega)=\sum_{l\in\partial\Lambda}\prod_{i\in {\cal P}_l}\left[\frac{\epsilon_{i}\,g/K}{\epsilon_{i}^{2}-(\omega/2)^{2}}\right]^{2}J_b(\omega).
\eea
Similarly as in the XY model, we obtain
\bea
\label{gammaIsing}
\gamma_{\rm Ising}(\omega)&\equiv& -\lim_{L\rightarrow \infty}\frac{1}{L}\overline{\log\left[\frac{\Gamma_{0}(\omega)}{J_b(\omega)}\right]}\nn\\
&=& -\left[\log\left(\frac{g}{K}\right)+\min_{x\in [0,1]}h_{\omega}(x)\right],
\eea
where the function $h_{\omega}(x)$ is now given by
\bea
\label{targetfunctionIsing}
h_{\omega}(x)=\frac{1}{2x}\log\left[K \int^{1}_{-1}\left|\frac{\epsilon}{\epsilon^2-(\omega/2)^2}\right|^{2x} \frac{d\epsilon}{2} \right].
\eea
Delocalization at energy $\omega$ occurs when $\gamma_{\rm Ising}(\omega)=0$. We find again that in the paramagnetic phase ($g<g_{c}$) no finite $\omega>0 $ satisfies this condition. This implies that at finite energies, excitations are localized in the disordered phase, similarly as in the XY model. 

For $0< \omega \ll 1$, by expanding $h_\omega$ around $\omega=0$, one finds
\bea
h_{\omega}(x) 
&=& \frac{1}{2x}\log\left\{ K \left[\frac{1}{1-2x}
-\frac{2x}{1+2x}\left(\frac{\omega}{2}\right)^2+o(\omega^2) \right]\right\}\nn\\
&=&  \frac{1}{2x}\log \frac{K}{1-2x}-\frac{1-2x}{1+2x}\left(\frac{\omega}{2}\right)^{2}+o(\omega^2).
\eea
A correction of order $\omega^{1-2x}$ which one might expect at first sight, can be shown to cancel. 

Close to the critical point, $g=g_c$,
we finally obtain the frequency-dependent decay rate 
\bea
\label{Isingdecayrate}
\gamma_{\rm Ising}(\omega)&=&-\left[\log\left(\frac{g}{K}\right)+\min_{x\in[0,1/2]}h_{\omega}(x)\right]\\
&=&\delta 
+\frac{1-2x_c}{1+2x_c}\left(\frac{\omega}{2}\right)^2+o(\omega^2),\nn
\eea
where $x_c$ is determined by $dh_{\omega=0}/dx(x=x_c)=0$, and is
related to $g_c$ by the condition for criticality,
\bea
\label{hw0_largeK}
h_{\omega=0}(x_c) &=&\frac{1}{1-2x_c}= -\log \left(\frac{g_c}{K}\right).
\eea
However, as we will see in section \ref{ss:rareregionsIsing} below, this leading order result is substantially modified by non-perturbative corrections.

\subsection{Critical point of the Ising model}
Note that Eq.~(\ref{hw0_largeK}) leads to the same expressions for $g_c$ and $x_c$ as derived above in Eq.~(\ref{SItransition}) for the XY case. However, as we will argue in the following subsection the leading asymptotic expression for $g_c$ in Eq.~(\ref{naivegc}) is  {\em not} renormalized by higher order corrections, unlike in the XY case. This implies that on the Cayley tree,  
\bea
\label{gcs}
g_{c,{\rm Ising}}\approx \frac{2}{e}g_{c,{\rm XY}}\approx \frac{1}{e \log K},\quad {\rm for}\quad K\gg 1. 
\eea
In other words, the Ising model orders at {\em weaker} exchange than the XY model, 
given an equal strength of random field disorder. 
This essentially reflects the fact that propagation of order in the 
Ising model is stronger at $\omega=0$, as resonances at 
low energy are not softened by self-energy effects, unlike in the XY model.

We conjecture that the ratio $g_{c,{\rm Ising}}/g_{c,{\rm XY}} \approx 2/e$ yields a good estimate for the ratio of critical points in the limit of high dimensional lattices.

\subsection{Effects from rare regions on spatial decay rates in the Ising paramagnet}
\label{ss:rareregionsIsing}
Note that so far our analysis for frequency dependent decay rates in the Ising model was restricted to the leading order in exchange. 
This yielded corrections $\gamma(\omega)-\gamma(0)\approx A \omega^2$ for both the Ising model and the XY model, 
with a coefficient $A$, which is only slightly bigger in the Ising case. 

However, from the 1-dimensional Ising chain studied in Sec.~\ref{s:1d}, it is clear that for the Ising model we should expect strong corrections to this 
quadratic behavior, because of the special role played by $\omega=0$ and the analogue of the rare stretches that we identified in 1d.
The effect of subleading corrections in the Ising model is indeed quite different from the XY case.
As we argued in Sec.~\ref{ss:resonances}, and as also appears clearly from the explicit calculations in the 1d chain, 
self-energy corrections are essentially absent in the limit $\omega\to 0$, while they do appear at finite $\omega$. 
On the other hand, as we discussed above for the XY model, the
regularizing self-energy corrections along the dominant delocalizing path at $\omega=0$ are responsible 
for the leading corrections to the decay rate, both in free fermions and in the XY model.~\cite{Abou-Chacra, Victor2013}
Since such self-energy corrections are absent along 1d Ising chains, 
we expect that the first corrections to $g_{c,{\rm Ising}}$ will be {\em subleading} in the large $K$ limit. Hence we expect that Eq.~(\ref{gcs}) captures the correct asymptotics for the critical point of the Ising model.  

\begin{figure}
\includegraphics[width=3.0in]{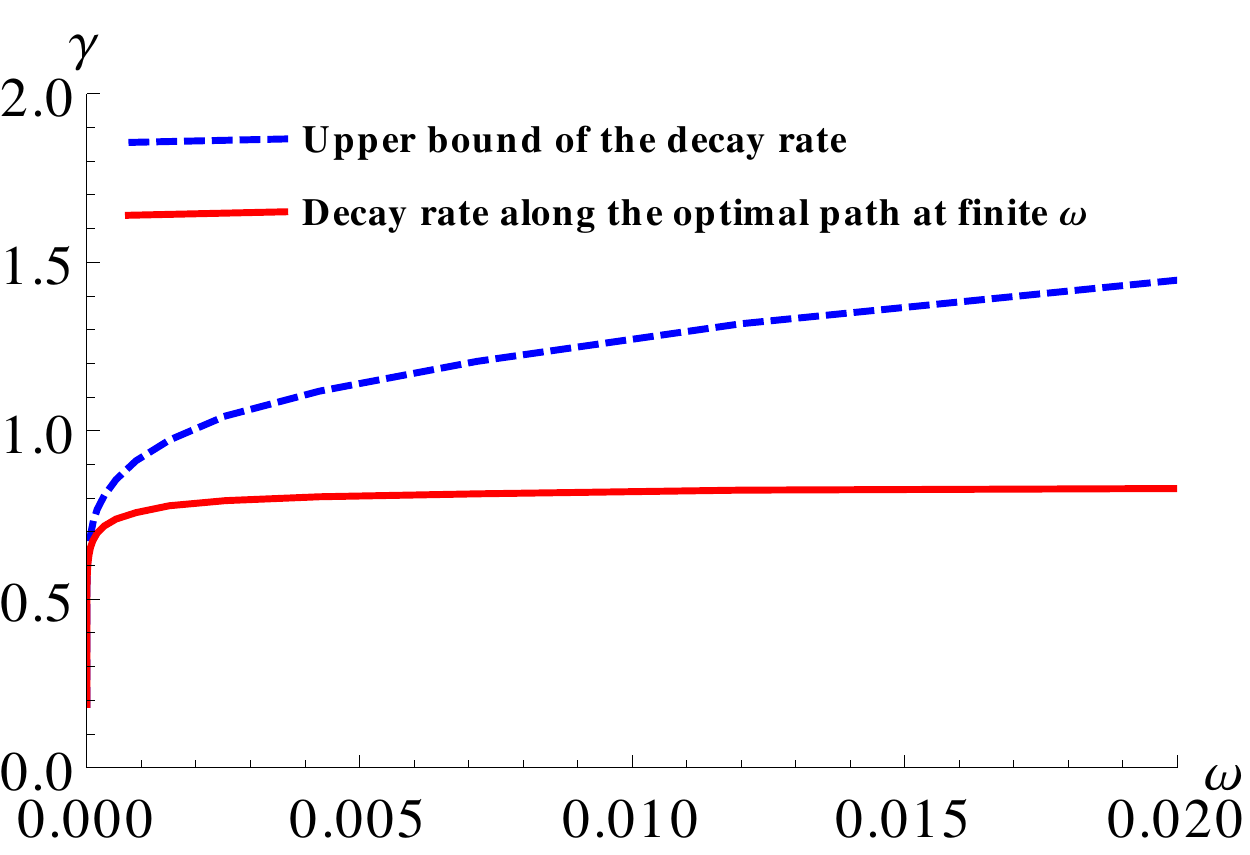}
\vspace{.2cm}
\caption{(Color online) Spatial decay rate $\gamma$ as a function of frequency $\omega$ for the Ising model 
 at the transition point for a tree branching ratio $K=2$. $\gamma$ is strictly positive for $\omega>0$. 
The red solid line shows the decay rate along a rare path with a biased, $\omega$-dependent disorder distribution (\ref{rhoomega}). 
The blue dashed line shows the upper bound obtained by restricting the propagation to a path ${\cal P}_0$ with distribution (\ref{poptimal}) which is optimal for propagation $\omega=0$.
At low energy, the two estimates nearly coincide and exhibit the activated scaling $1/|\log(\omega)|^\psi$; cf. Fig~\ref{f:tightnessoftheupperbound}.} 
\label{f:Decayratesalongthebestpathatfiniteenergy}
\end{figure}

\begin{figure}
\includegraphics[width=3.0in]{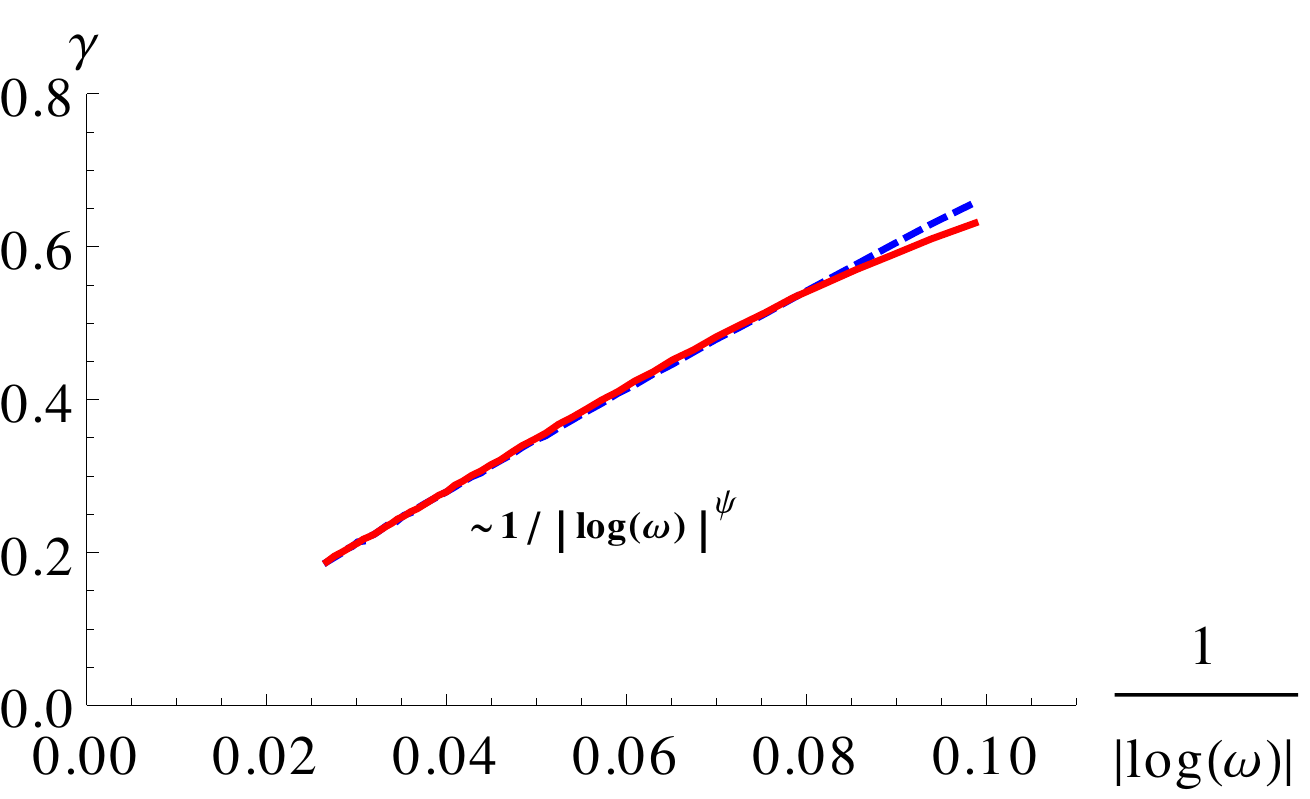}
\vspace{.2cm}
\caption{(Color online) The same data as in Fig.~\ref{f:Decayratesalongthebestpathatfiniteenergy}, plotted against $1/|\log(\omega)|$ at low energy.
The two estimates nearly coincide and exhibit the activated scaling $1/|\log(\omega)|^\psi$ with $\psi=0.98\pm 0.01$.}
\label{f:tightnessoftheupperbound}
\end{figure}

To assess the effect of rare regions on excitations at finite energy, we need to study the spatial decay rate of finite energy excitations.
One expects that such excitations still decay preferentially along one preferred path, which optimizes the Lyapunov exponent at that given frequency. However, this path in general depends on the frequency, and thus is not necessarily identical to the path ${\cal P}_0$ which optimizes propagation at $\omega=0$.  
Nevertheless, by restricting the analysis of propagation to ${\cal P}_0$ we obtain an upper bound on $\gamma_{\rm Ising}(\omega)$. 
This bound turns out to be rather tight at very low $\omega$, 
as we confirmed by comparing it with a calculation where the paths are optimized for each frequency individually, cf. Fig~\ref{f:Decayratesalongthebestpathatfiniteenergy}.

\subsubsection{Biased energy distribution on optimal paths ($\omega=0$)}
Let us first point out some interesting properties of the optimal path ${\cal P}_0$.
 It is physically intuitive that the path at $\omega=0$ will have a non-uniform distribution of 
on-site energies, peaked around $\epsilon_i=0$. More precisely, by exploiting the mapping to directed polymers, 
and using the exact solution by Derrida and Spohn,~\cite{directedpolymer} one can show that the onsite energies along the optimal path are distributed according to the density
\bea
\label{poptimal}
\rho(\epsilon)=\frac{1-2x_c}{\epsilon^{2x_c}},\quad {\rm for}\quad \epsilon\in [0,1].
\eea
This follows from the observation that the locators $1/\epsilon$ take the role of local Boltzmann weights of the directed polymer. In the frozen phase of the polymer, where a single path dominates, the distribution of visited $\epsilon$'s is the same as  that in a high temperature phase with an effective temperature $T_{\rm eff}= T/x_c$, i.e., with corresponding modified Boltzmann weights $1/e^{x_c}$.~\cite{directedpolymer} Hereby, $T_{\rm eff}$ is the temperature where the freezing transition occurs, which is the last temperature where an annealed calculation of the partition function can be used. From the latter it follows that sites with given locators are visited by the polymer with a probability proportional to their modified Boltzmann weight, which implies Eq.~(\ref{poptimal}).~\cite{VictorHint} Using this distribution of onsite disorder along the optimal path, the criterion Eq.~(\ref{criticalconditionfor1dIsing}) 
for the critical point on this path correctly yields the critical exchange $g_{c,{\rm Ising}}$ of Eq.~(\ref{gcs}), as it must be.

It is instructive to obtain this result in a different way, which is generalizable to finite frequencies. 
Anticipating that delocalization occurs along the most favorable path, we should find rare paths of probability $P \approx 1/K^L$ with optimally biased distribution of onsite energies, such that the Lyapunov exponent is maximized. Let $\rho(\epsilon)$ be that distribution. The probability $P_\rho$ of finding a path of length $L$ with such a biased distribution is given by the relative decrease in entropy as compared to the uniform distribution found on typical paths,
\bea
\ln(P_\rho)= L \int_0^1 d\epsilon \rho(\epsilon) \ln[\rho(\epsilon)]. 
\eea
Neglecting, as previously, the small corrections from sites adjacent to the path~\cite{footnotelast} the Lyapunov exponent at $\omega=0$ is simply 
\bea
\gamma_\rho(\omega=0)= \int_0^1 d\epsilon \rho(\epsilon) \ln(J/\epsilon).
\eea
Maximizing this with respect to $\rho$ under the constraint $\ln(P_\rho)/L = -\ln(K)$ yields back the power law distribution (\ref{poptimal}).

\subsubsection{Optimal paths at finite $\omega$ - activated scaling}
  
To optimize the propagation path for finite energy excitations  is more complex, since at finite $\omega$ the Lyapunov exponent along a path is not simply the logarithm of a  product of factors on each site, but the eigenvalue of a product of non-commuting matrices. This implies that the Lyapunov exponent 
not only depends on the distribution of energies along the path, but also on their sequence and correlations. We will neglect this fact and optimize with respect to distributions $\rho(\epsilon)$, which are uncorrelated from site to site. However, since the Lyapunov exponent cannot be expressed simply as a functional of $\rho$, we estimate $\gamma(\omega)$ by the {\em Ansatz}, generalizing (\ref{poptimal}) to 
\bea
\label{rhoomega}
\rho_\omega(\epsilon) = \frac{A}{|\omega-\epsilon|^\sigma}.
\eea
Hereby $A$ and $\sigma$ have to be determined by the normalization condition and the requirement that a path with this atypical distribution can be found with finite probability among the $K^L$ paths, i.e., $\ln(P_\rho)/L = -\ln(K)$.
The resulting decay rate $\gamma(\omega)$ is plotted in Fig.~\ref{f:Decayratesalongthebestpathatfiniteenergy} for $K=2$. 
We show the data together with the upper bound obtained from the decay at finite $\omega$ on the path ${\cal P}_0$ with energy distribution (\ref{poptimal}). 
At very low energies the two estimates nearly coincide, cf. Fig.~\ref{f:Decayratesalongthebestpathatfiniteenergy}.

To discuss the off-critical behavior at low frequencies, we therefore restrict ourselves to the fixed path ${\cal P}_0$ which is optimized for $\omega=0$. Using Eqs.~(\ref{Legendre},\ref{expo}) for the Lyapunov exponent along this fixed path, and the relation (\ref{xcXY}) we obtain the equation
\bea
\log \left(1-\frac{\alpha}{1-2x_c}\right)+ \alpha \left(\frac{1}{1-2x_c}+\delta\right)=0
\eea
for the exponent $\alpha$, which governs the non-analytic correction 
\bea
\label{gammaomegaCayley}
\gamma(\omega)-\gamma(0)=O(\omega^\alpha).
\eea
As previously, the distance from criticality is defined as $\delta =\log(J_c/J)$. 
Close to the transition point $\alpha \simeq 2(1-2x_c)^2 \delta$, which vanishes at criticality. 
This suggests that the non-analytic correction vanishes more slowly than any power law as $\omega\to 0$.
 Therefore, we expect the critical scaling to be {\em activated} with  
\bea
\label{loglaw}
\gamma_{\rm Ising}(\omega)=\xi_{\rm Ising}^{-1}(\omega) \propto \frac{1}{|\log(\omega)|^{\psi}}.
\eea
This is in agreement with the empiric observation of infinite randomness fixed points in Ising models on Erd\"os-R\'enyi graphs.~\cite{KovacsIgloi3d} 

\begin{figure}
\includegraphics[width=3.0in]{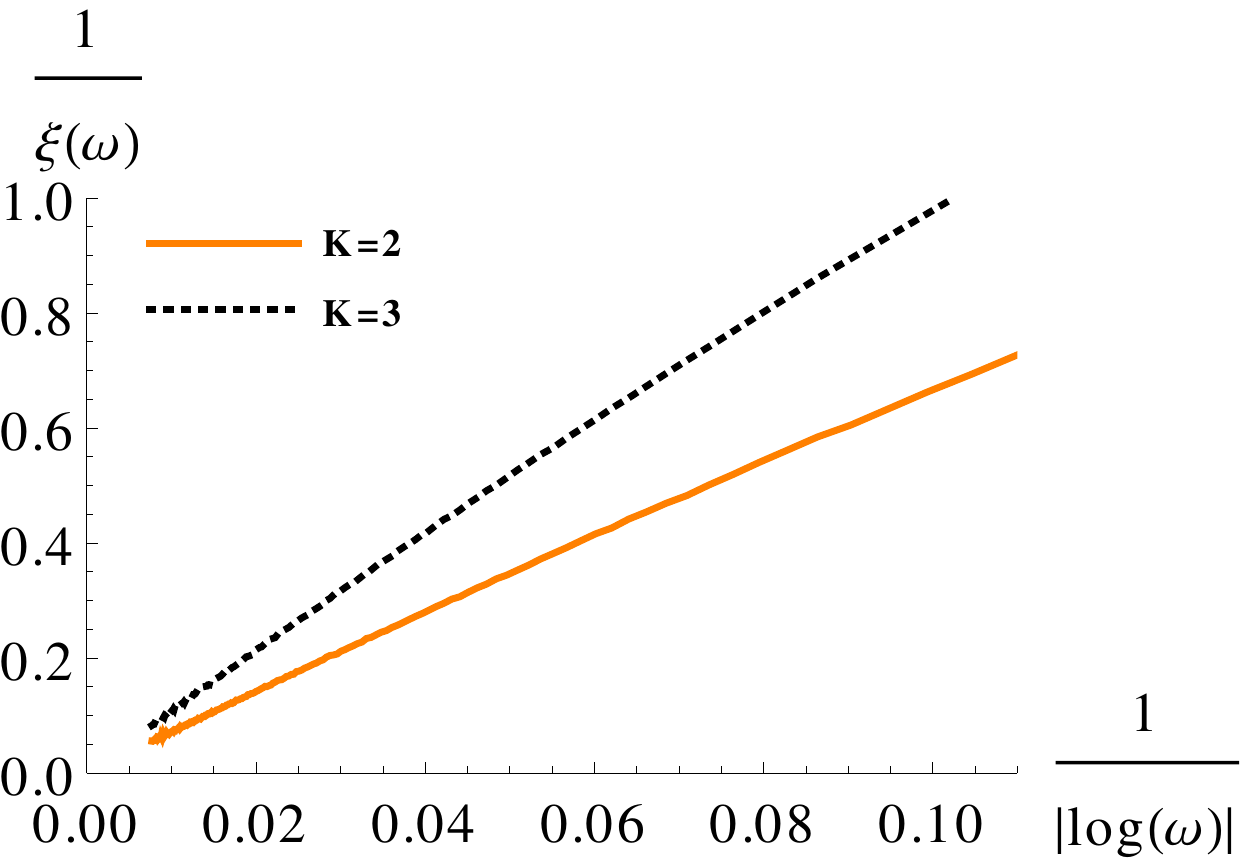}
\vspace{.2cm}
\caption{
Inverse of the localization length in the Ising model as a function of frequency
$\omega$. It is evaluated  at the critical point for
connectivity $K=2,3$ along the path that optimizes propagation at $\omega=0$. 
 The numerical data is fitted to 
a logarithmic dependence $\xi_{\rm Ising}^{-1}(\omega) \sim \frac{1}{|\log(\omega)|^{\psi}}$, 
yielding values $\psi\simeq 0.98\pm 0.01$ consistent with the expected $\psi=1$, independently of $K$. 
 }
\label{f:Decayratesalongthebestpath}
\end{figure}

We fit the expected law (\ref{loglaw}) to the data obtained for $\gamma(\omega)$ at low $\omega$, along optimized paths with biased distributions of the form (\ref{rhoomega}). 
This yields an  exponent $\psi\approx 0.98\pm 0.01$. This exponent is consistent with the value $\psi=1$, which we obtained rigorously for $1d$ chains.  
In Fig.~\ref{f:Decayratesalongthebestpath} 
we show data for $\gamma(\omega)$ evaluated on ${\cal P}_0$, illustrating that the essentially linear behavior in $1/\log(\omega)$ 
is essentially independent of the branching ratio $K$. 

The exponent $\psi=1$ is expected from scaling considerations. Let us suppose that close to criticality, at low $\omega$, $\gamma$ assumes a scaling behavior. From Eq.~(\ref{gammaomegaCayley}) we expect that $\gamma= \delta +{\rm const\cdot\,}\omega^\alpha$ with $\alpha\propto \delta$. This suggests, in analogy to Eq.~(\ref{scalingxi}), 
that there should be a scaling limit, $\delta \rightarrow 0$ and $\omega \rightarrow 0$, whereby $\delta |\log(\omega)|$ is kept constant, 
and $\gamma_{\rm Ising}(\omega)$ tends the following scaling form
\bea
\label{scalingxibestpath}
\gamma_{\rm Ising}(\omega) \rightarrow \delta f(\delta |\log \omega|).
\eea
The scaling function $f(x)$ should behave as $f(x)\rightarrow 1 + a\exp (-bx)$  
for $x\gg 1$, with positive constants $a$ and $b$.
  
To obtain a finite limit when $\delta \rightarrow 0$, 
$f(x)$ must behave as $\sim 1/x$ when $ x \rightarrow 0$, from which one infers the  
\bea
\gamma_{\rm Ising}(\omega;\delta=0) \sim 1/|\log \omega|.
\eea

\subsubsection{Ising vs. XY model - comparison with finite $d$}
On the Cayley tree the decrease of the localization length with increasing $\omega$ in the critical Ising model is thus much faster than in the critical XY model. This matches with what is known in finite dimensions from various numerical approaches. Indeed, the  Ising model in strong random field disorder is known to be governed by infinite randomness fixed points, both in low dimensions~\cite{Motrunich,KovacsIgloi2d} and in the infinite-dimensional limit of random Erd\"os-R\'enyi graphs.~\cite{KovacsIgloi3d} The infinite randomness fixed points exhibit activated scaling, which implies that at criticality the inverse localization length grows only logarithmically with energy, $\xi_{\rm typ} = \gamma^{-1} \sim |\log(\omega)|^\psi$. In contrast, XY models in dimensions $d>1$ were found to exhibit power law scaling $\xi_{\rm typ}= \gamma^{-1} \sim \omega^{-1/z}$.~\cite{VojtaReview} As we will discuss next, however, our calculation for the XY model on the Cayley tree does not apply at finite frequencies at criticality, and thus does not allow us to infer the dynamical exponent $z$ in the limit of infinite dimensions.

\subsection{Range of applicability of the results close to criticality}
At this point we should recall the limits of validity of the locator expansion on which the above calculations are based. Far from criticality, 
the localization properties of intensive excitations (i.e., excitations at $T=0$) 
are well described by low orders of perturbation theory, once the physics of resonances has been taken into account. 
However, close to criticality, one has to be more careful, as the convergence of the perturbation expansion is much slower, 
which requires an analysis of the contributions from significantly higher orders in perturbation theory. 

This is well illustrated by the discussion of a simple condition  which the localization length as a function of $\omega$ must satisfy. 
Suppose a finite energy excitation of energy $\omega$ splits effectively into $n$ independently propagating excitations of energy $\omega_n$, 
which do not interfere with each other at large distances. This is one possibility of a high order process, 
which goes beyond the calculations presented above. 
Indeed it only occurs at orders $O(J^{nL})$ of perturbation theory for correlation functions at distances $L$.
Supposing that excitations of energy $\omega$ decay in space with rate $\gamma(\omega)$, $n$ independently propagating packets of energy will decay as a whole like $\exp[-\sum_{i=1}^n\gamma(\omega_i) L]$. For large distances this propagation mode yields a lower bound for the decay rate of an excitation of energy $\omega=\sum_{i=1}^n\omega_i$. Applied to the spin models considered here, we thus obtain the consistency condition
\bea
\label{elasticcondition}
\gamma(\omega)\leq \sum_{i=1}^n \gamma(\omega_i), \quad \omega=\sum_{i=1}^n\omega_i, 
\eea
for all {\em odd} integers $n>1$ (since a single spin flip can only split into an odd number of individually propagating excitations, due to parity conservation).  

The result (\ref{loglaw}) for $\gamma(\omega)$ in the Ising model satisfies this inequality by large in a broad range of low frequencies, at any distance $\delta$ from criticality. We thus do not expect that higher order corrections modify the results obtained in previous sections for low frequencies, even very close to criticality. In particular this seems to preclude the possibility that the Ising transition coincides with the closing of a mobility gap in the paramagnet. Instead our result suggests that the Ising order emerges by a delocalization phenomenon at $\omega=0$, while all low energy excitations (in a fixed finite range) are still localized in the paramagnet.

In contrast, the result for hard core bosons (XY model) has a more restricted range of applicability. A behavior of the form 
$\gamma(\omega)=\delta+C\omega^2$ as predicted by Eq.~(\ref{gammaXY}), is certainly valid without serious restrictions for non-interacting fermions. However, for interacting hard core bosons
 it is consistent with Eq.~(\ref{elasticcondition}) only in the low frequency range
\bea
\label{range}
0\leq \omega\lesssim \sqrt{\delta/C}, 
\eea
which vanishes upon approaching the transition, $\delta \to 0$. 
This indicates that much higher orders processes must be studied to describe the localization properties of {\em finite} energy excitations very close to criticality in the XY model. 

Nevertheless, we conjecture that the parabolic increase of the decay rate $\gamma$ with frequency does hold in the range (\ref{range}). This is almost certain for large enough $\delta$, due to the fast spatial decay of excitations in that regime. However, a rigorous proof for small $\delta$ would require an involved analysis of the propagation of several lumps of energy $\omega_n$ and their scattering from each other, to ensure that the latter does not significantly enhance the delocalization tendency of the considered excitation. While we cannot exclude this possibility, it does not seem plausible to us at low enough energies, due to the lack of phase space for such scatterings. 

This then suggests that at any finite $\delta$ within the paramagnet $\gamma(\omega=0)$ is locally minimal at $\omega=0$. 
However, due to the restricted range (\ref{range}) of applicability of our results for the XY model, 
we cannot firmly exclude the presence of many body mobility edges at some $\omega_c(\delta)>\sqrt{\delta/C}$, 
which nevertheless might tend to zero as $\delta\to 0$. 
The possibility of a finite energy mobility edge in similar spin models has been suggested by numerics on small random graphs,~\cite{Cuevas} which is however subject to several caveats.~\cite{positivemagnetoresistance}

Finally, we should mention here that such mobility edges do exist almost trivially if the distribution of  onsite energies is not flat at the chemical potential, but has a substantial slope. 
This allows for better hybridization, that is, enhanced propagation of excitations at higher energies. 
It is then easy to show from the formulas in the previous subsections, that excitations at high enough energy are delocalized while lowest energy excitations are still localized. The non-trivial content of our results discussed above consists in predicting that in the paramagnetic phase, close enough to $\omega=0$ the localization length always decreases with increasing  $\omega$, if the density of bare disorder is uniform in a sufficiently wide interval.

\section{Nature of the ordering transition and fractality}
\label{s:RSB}
As mentioned in Sec.~\ref{ss:dirpol}, the value $x_c$ which minimizes the functions $f(x)=f_{\omega=0}(x/2)$ (Ising) 
and $f_{\omega=0}^\Delta(x/2)$ (XY) determining the static surface susceptibility, actually contains physical meaning. If at the critical point, where $f(x_c)=0$ one finds $x_c<1$, the response to a symmetry breaking field applied at the surface is dominated by a finite number of different paths. In contrast, for $x_c=1$ an infinite number of paths contributes to the response in the thermodynamic limit. More precisely, in the thermodynamic limit $1-x_c$ is known to be equal to the disorder average of the following double sum 
over boundary sites~\cite{directedpolymer} 
\bea
\label{IPR}
1-x_c = \overline{\sum_{l, l'} {}^{(q)}\,w_{l}w_{l'}}\,,
\eea
where $w_{l}$ is the relative weight of the path from the root to site $l$ in the total surface susceptibility, 
\bea
w_{l} = \frac{G_{l,0}(\omega=0)}{\sum_{l'}G_{l',0}(\omega=0)}.
\eea
The sum $\sum^{(q)}(...)$ in Eq.~(\ref{IPR}) restricts the sites $l,l'$ to pairs whose first common ancestor is not too close to the root, but at least a finite fraction $q$ of the diameter $L$ from it. 
In analogy to spin glass problems, $q$ is called the overlap between the paths leading to $l$ and $l'$, respectively. 
Remarkably, the result (\ref{IPR}) was shown to be {\em independent} of the value chosen for the overlap $1>q>0$. 
This is possible because in the thermodynamic limit the relevant contributions are due to pairs of sites with overlap $q\to 1$. 
Those belong to small clusters of sites, different clusters being very far from each other, with mutual overlaps $q\to 0$ between paths ending in different clusters.   
 
The above thus allows one to rewrite $1-x_c$ more suggestively as an inverse participation ratio of cluster contributions,
\bea
\label{IPR2}
1-x_c =\overline{ \sum_{\cal C}w_{\cal C}^2},
\eea
where  $w_{\cal C}=\sum_{i\in {\cal C}}$ is the weight of the cluster $\cal C$. 

The Ising and XY models crucially differ, as $1-x_c$ is finite for the first and vanishes for the latter. Indeed, we have argued that for the Ising model the minimum occurs at $x_c<1$, implying that the ordered state develops on a very sparse subgraph of the lattice, which essentially corresponds to a finite number $\sim 1/(1-x_c)$ of narrowly clustered paths. In contrast, in the XY model   
 we have argued that self-energy corrections bring the minimizing value of $x$ to $x_c=1$. This implies that the incipient transverse order establishes on a less sparse subgraph, since infinitely many clusters of paths are implicated in establishing it. Nevertheless the forming condensate is still very inhomogeneous. Indeed, the number of paths contributing to the susceptibility (as defined by the inverse of (\ref{IPR})), even though infinite in the thermodynamic limit, is still much smaller than the total number of paths ($\sim K^L$). In fact, it does not even grow exponentially with the diameter of the tree, the configurational entropy of paths being proportional to $df^\Delta_{\omega=0}/dx(x=x_c/2)=0$, which vanishes, because $x_c=1$ is a local minimum of $f^\Delta_\omega=0$.    
  
The fact that special paths tend to dominate the propagation in disordered quantum systems was already anticipated by Anderson 
in his seminal paper on single particle localization, whereby he assumed statistical independence of different paths.~\cite{Anderson}
While the latter turns out to be almost literally true on the Cayley tree,~\cite{Abou-Chacra,BiroliTarzia} propagation paths 
in finite dimensions are much more correlated.~\cite{ThoulessReview1970} Nevertheless, one should probably consider the sparsity (and multifractality) of 
critical wavefunctions in finite-dimensional Anderson models~\cite{MirlinEvers} as a remnant of the much stronger dominance by a few single paths on the Cayley tree.

Such fractality should not be specific to non-interacting problems. In fact, it is natural to expect multifractality also at many body ordering transitions in the presence of strong enough quenched randomness. 
A known example of such a phenomenon is the phase transition in classical disordered Pott's ferromagnets in 2d.~\cite{Ludwig, ChatelainBerche}
A real space RG study~\cite{Motrunich} of the random transverse field Ising model in 2d also suggested that order sets in on a percolating cluster of fractal dimension $d_f\approx 1$. 

In order to define more precisely the concept of (multi)fractality of an emerging long range order,
we propose as a natural object of study the spatially resolved response to a transverse field $\psi_i = d\langle s^x_i\rangle/dh_x$.~\cite{footnotefractality}   
At the critical point,  $\psi_i$ is expected to behave similarly to a critical wavefunction at a single particle Anderson transition. 
In particular, we expect multifractality in the form
\bea
\frac{\sum_i |\psi_i|^{2q}}{(\sum_i |\psi_i|^2)^q} \propto L^{-d_q(q-1)}
\eea
with non-trivial fractal dimensions $d_q<d$, $L$ being the linear size of the $d$-dimensional system.

Our result on the Bethe lattice, namely that the sparsity of the critical Ising condensate is more pronounced than that of the XY condensate, 
leads us to conjecture that on finite dimensional lattices the above defined fractal dimensions will be larger for XY models than for Ising models,
\bea
d_q^{\rm Ising}\leq d_q^{\rm XY},
\eea
with strict inequalities at least in some range of $q$-values. 
It would be interesting to check this prediction numerically.

\section{Conclusion}
\label{s:discussion}

The two spin models considered in this paper - the random transverse field Ising model and the XY model (hard core bosons) in disorder potentials - turn out to behave rather differently. They have in common, that in the paramagnetic phase, with a uniform, unbiased disorder distribution, at low $\omega$ the localization length tends to decrease with increasing $\omega$. However, this increase is much more dramatic for Ising models than for hard core bosons. This is due to the  Ising symmetry which protects resonances at  low frequencies, while self-energy corrections regularize resonances in the XY model, similarly as in free fermion models. This difference is at the root of the activated critical scaling at the Ising transitions, whereas we expect standard power law scaling in the XY model. 

The strong decrease of the localization length with increasing energy in the Ising case suggests that the ordering transition can be analyzed without resorting to very high orders in perturbation theory. Under this assumption we found that order establishes due to a delocalization phenomenon 
that is initiated at $\omega=0$, while excitations at small but finite $\omega$ are still rather strongly localized. For uniform disorder we did not find evidence that there are any high energy (but intensive) excitations which are already delocalized at this point. However, we cannot exclude it, as such delocalization phenomena might only show up at high energies, in much higher orders of perturbation theory than we analyzed here.
It would be interesting to study the evolution of localization or diffusion properties of finite energy excitations just slightly within the Ising ordered phase.

On a Cayley tree, the magnetic order is found to appear first on an extremely sparse subgraph, much sparser than for the equivalent ordering transition in the XY model. We expect that this difference also persists in finite dimensions, where we conjecture Ising condensates to emerge on percolating structures with lower fractal dimensions than their XY counterparts.
 
In equivalent random field disorder we found that XY models order only at stronger exchange couplings than Ising systems, because resonances  at low energies are regularized, unlike in the Ising models. 
In the XY case, however, the critical regime cannot faithfully be studied at the orders of perturbation theory considered here, as the controllable frequency range shrinks to zero upon approaching criticality. It remains an interesting open problem to describe the dynamics of finite energy excitations in a (nearly) critical XY system. In particular, we expect the diffusion of an initial energy packet to be significantly more complex than in an Ising model.
The differences between hard core bosons and fermions with respect to such delocalization transitions also remain an interesting subject for further study.

\begin{acknowledgments}
We thank V. Bapst, M. Fabrizio, V. Gurarie, V. Kravtsov, C. Laumann, P. Le Doussal, M. Tarzia and F. Zamponi for useful discussions.
MM thanks KITP Santa Barbara for hospitality, while part of this work was accomplished. This
research was supported in part by NSF-KITP-12-184.
\end{acknowledgments}

\appendix

\section {Ising chain with three spins }

As a non-trivial check of the result of the locator expansion Eq.~(\ref{Isingmatrixelement}), we calculate the matrix element $\langle \textrm{GS}|\sigma^{x}_{l}|E_{0}\rangle$ for a three spin chain by standard perturbation theory.
Without loss of generality we may suppose that $\epsilon_{i}>0$. Then $|\textrm{GS}^{(0)}\rangle=|\uparrow\uparrow\uparrow\rangle$ and $|E^{(0)}_{0}\rangle=\sigma^{-}_{0}|\textrm{GS}^{(0)}\rangle=|\downarrow\uparrow\uparrow\rangle$.

We now evaluate $\langle E_{0}|\sigma^{x}_2|\textrm{GS}\rangle$ by standard perturbation theory.
The Hamiltonian is
\bea
H=-\sum^2_{i=0}\epsilon_{i}\sigma^{z}_{i}-t\sum^{1}_{i=0}\sigma^{x}_{i}\sigma^{x}_{i+1}.
\eea
We treat the exchange  term $H_{I}\equiv-t\sum^{1}_{i=0}\sigma^{x}_{i}\sigma^{x}_{i+1}$ as a perturbation and denote the matrix element $H^{kl}_{I}\equiv\langle E^{(0)}_k|H_{I}|E^{(0)}_l\rangle$.
The perturbed eigenstates adiabatically connected to $|E^{(0)}_{\textrm{GS}}\rangle$ and to the excited state $|E^{(0)}_{0}\rangle$ are 
\bea
|\textrm{GS}\rangle&=&|E^{(0)}_{\textrm{GS}}\rangle-t\sum_{k\neq \textrm{GS}}\frac{H^{\textrm{GS}k}_{I}}{E^{(0)}_{\textrm{GS}}-E^{(0)}_{k}}|E^{(0)}_{k}\rangle \nn\\
&&+t^{2}\sum_{k\neq\textrm{GS}}\sum_{l\neq\textrm{GS}}\frac{H^{kl}_{I}H^{\textrm{GS}l}_{I}}{(E^{(0)}_{\textrm{GS}}-E^{(0)}_{k})(E^{(0)}_{\textrm{GS}}-E^{(0)}_{l})}|E^{(0)}_{k}\rangle\nn\\
&&-\frac{t^{2}}{2}\sum_{k\neq\textrm{GS}}\frac{|H^{k\textrm{GS}}_{I}|^2}{(E^{(0)}_{\textrm{GS}}-E^{(0)}_{k})^2}|E^{(0)}_{\textrm{GS}}\rangle +O(t^{3})\nn\\
&=&|\uparrow\uparrow\uparrow\rangle-t\left(\frac{1}{-2\epsilon_{0}-2\epsilon_{1}}|\downarrow\downarrow\uparrow\rangle+\frac{1}{-2\epsilon_{2}-2\epsilon_{1}}|\uparrow\downarrow\downarrow\rangle\right)\nn\\
&&+t^{2}\left[\left( \frac{1}{(-2\epsilon_{0}-2\epsilon_{2})(-2\epsilon_{0}-2\epsilon_{1})}\right.\right.\nn\\
&&\left.\left.+\frac{1}{(-2\epsilon_{0}-2\epsilon_{2})(-2\epsilon_{1}-2\epsilon_{2})}\right)|\downarrow\uparrow\downarrow\rangle \right.\nn\\ &&\left.-\frac{1}{2}\left(\frac{1}{(2\epsilon_{0}+2\epsilon_{1})^{2}}+\frac{1}{(2\epsilon_{2}+2\epsilon_{1})^{2}}\right)|\uparrow\uparrow\uparrow\rangle\right]+O(t^{3})\nn\\
&\equiv& E|\uparrow\uparrow\uparrow\rangle+F|\downarrow\downarrow\uparrow\rangle+G|\uparrow\downarrow\downarrow\rangle+H|\downarrow\uparrow\downarrow\rangle\nn\\
&&+O(t^{3}),
\eea
where
\bea
E&=&1-\frac{t^2}{2}\left(\frac{1}{(2\epsilon_{0}+2\epsilon_{1})^{2}}+\frac{1}{(2\epsilon_{2}+2\epsilon_{1})^{2}}\right),\nn\\
F&=& \frac{t}{2\epsilon_{0}+2\epsilon_{1}},\nn\\
G&=& \frac{t}{2\epsilon_{2}+2\epsilon_{1}},\\
H&=&\frac{t^{2}}{2\epsilon_{0}+2\epsilon_{2}}\left( \frac{1}{2\epsilon_{0}+2\epsilon_{1}}+\frac{1}{2\epsilon_{1}+2\epsilon_{2}}\right). \nn
\eea
For the excited state $|E_{0}\rangle$ we find similarly
\bea
|E_{0}\rangle&=&|E^{(0)}_{0}\rangle-t\sum_{k\neq 0}\frac{H^{0k}_{I}}{E^{(0)}_{0}-E^{(0)}_{k}}|E^{(0)}_{k}\rangle \nn\\
&&+t^{2}\sum_{k\neq0}\sum_{l\neq0}\frac{H^{kl}_{I}H^{0l}_{I}}{(E^{(0)}_{0}-E^{(0)}_{k})(E^{(0)}_{0}-E^{(0)}_{l})}|E^{(0)}_{k}\rangle\nn\\&&-\frac{t^2}{2}\sum_{k\neq0}\frac{|H^{k0}_{I}|^2}{(E^{(0)}_{0}-E^{(0)}_{k})^2}|E^{(0)}_{0}\rangle+O(t^{3})\nn\\
&=&|\downarrow\uparrow\uparrow\rangle -t\left(\frac{1}{2\epsilon_{0}-2\epsilon_{1}}|\uparrow\downarrow\uparrow\rangle+\frac{1}{-2\epsilon_{2}-2\epsilon_{1}}|\downarrow\downarrow\downarrow\rangle\right)\nn\\
&&+t^{2}\left[\left( \frac{1}{(2\epsilon_{0}-2\epsilon_{2})(2\epsilon_{0}-2\epsilon_{1})} \right.\right.\nn\\ &&\left.\left.+\frac{1}{(2\epsilon_{0}-2\epsilon_{2})(-2\epsilon_{1}-2\epsilon_{2})}\right)|\uparrow\uparrow\downarrow\rangle \right.\nn\\ &&\left.-\frac{1}{2}\left(\frac{1}{(2\epsilon_{0}-2\epsilon_{1})^{2}}+\frac{1}{(2\epsilon_{1}+2\epsilon_{2})^{2}}\right)|\downarrow\uparrow\uparrow\rangle\right]+O(t^{3})\nn\\
&\equiv& A|\downarrow\uparrow\uparrow\rangle+B|\uparrow\downarrow\uparrow\rangle+C|\downarrow\downarrow\downarrow\rangle+D|\uparrow\uparrow\downarrow\rangle \nn\\
&&+O(t^{3}).
\eea
where
\bea
A&=&1-\frac{t^2}{2}\left(\frac{1}{(2\epsilon_{0}-2\epsilon_{1})^{2}}+\frac{1}{(2\epsilon_{1}+2\epsilon_{2})^{2}}\right),\nn\\
B&=&\frac{-t}{2\epsilon_{0}-2\epsilon_{1}},\nn\\
C&=& \frac{t}{2\epsilon_{2}+2\epsilon_{1}},\\
D&=& \frac{t^2}{2\epsilon_{0}-2\epsilon_{2}}
\left(\frac{1}{2\epsilon_{0}-2\epsilon_{1}}+\frac{1}{-2\epsilon_{1}-2\epsilon_{2}}\right).\nn
\eea
For the matrix element of the bath operator $\sigma^x_2$ between the excited and ground states we thus obtain
\bea
\langle E_{0}|\sigma^{x}_2|\textrm{GS}\rangle &=&AH+BG+CF+DE+O(t^{3})\nn\\
&=& H+BG+CF+D+O(t^{3}).
\eea
Combining the second term of $H$ with $BG$ and the second term of $D$ with $CF$, the resulting four terms can be factorized into the form
\bea
\langle E_{0}|\sigma^{x}_2|\textrm{GS}\rangle &=& \left(\frac{t}{2\epsilon_{1}-\omega}+\frac{t}{2\epsilon_{1}+\omega}\right)\times \nn\\
&&\left(\frac{t}{2\epsilon_{2}-\omega}+\frac{t}{2\epsilon_{2}+\omega}\right)+O(t^{3}).\quad
\eea
with $\omega=2\epsilon_{0}$ being the excitation energy.
This coincides precisely with the result of the locator expansion, Eq.~(\ref{Isingmatrixelement}).


\begin{thebibliography}{99}
\bibitem{fn2}
{Strictly speaking, there are no normalizable eigenmodes with exactly $\omega_n=0$. 
However, there are modes with energies that are exponentially small in the system size, 
and for those the value given in Eq.~(\ref{xi_Delta}) still applies.}
\bibitem{fn3}
{Close to criticality, or for Gaussian distributed $\log(\epsilon_i/J)$, we can consider $\log X_{\ell}$ in Eq.~(\ref{Xl}) as a Gaussian random variable with variance $\sigma^2$ and mean value $\delta$. This is suggested by the central limit theorem, and holds as long as we do not probe far tails of the distribution. 
The probability of 
$\log X_{\ell}$ can thus be estimated as
$p_{\ell}(\log X_{\ell}=a) \approx \exp\left[-\frac{(a-\ell \delta)^2}{2\ell \sigma^2}\right]$,
neglecting pre-exponential factors. Here
$\delta =\left|\overline{\log(\epsilon_i/J)}\right|, 
\sigma^2=\left(\overline{[\log(\epsilon_i)]^2}-\overline{\log(\epsilon_i)}^2\right).
$
In this case the frequency dependent change of the Lyapunov exponent results as
$\Delta\xi^{-1}(\omega)\sim \int\frac{d \ell}{\ell} p_{\ell}(\log X_{\ell}=\log \omega)\propto \exp\left[-|\log \omega| \frac{2\delta }{\sigma^2}\right] =\omega^{2\delta/\sigma^2}$,
which yields the exponent $\alpha=2\delta /\sigma^2$. This agrees with Eq.~(\ref{alpharesult}), using that for box distributed $\epsilon_i$ $\sigma=1$.}
\bibitem{fn4}
{Strictly speaking, in a finite sample there are no fermionic modes exactly at  $\omega=0$. Nevertheless, there are eigenmodes with  eigenvalues $\omega$ that are exponentially small in the system size, 
and their localization properties can be obtained without loss of accuracy by setting $\omega\to 0$.}
\bibitem{footnotelast} To a large extent their small effect can be incorporated in an effective distribution of the onsite energies.
\bibitem{footnotefractality} Alternatively, for the XY model, one may consider the macroscopically occupied eigenvector in the one-particle density matrix within the ordered phase, and study its limit as the critical point is approached.
\end{thebibliography}
\end{document}